\documentclass[conf]{new-aiaa}
\usepackage[utf8]{inputenc}

\usepackage{graphicx}
\usepackage{graphbox}
\graphicspath{{figures/}}
\usepackage{amsmath,amsfonts}
\usepackage[version=4]{mhchem}
\usepackage{siunitx}
\usepackage{longtable,tabularx}
\usepackage{tikz}
\usepackage{adjustbox}
\usepackage{multicol}
\usepackage{blindtext}
\usepackage{float}
\usepackage{tablefootnote}
\usepackage{color}
\usepackage{mathtools}
\usepackage{multirow}
\usepackage{hyperref}
\hypersetup{
pdftitle={From Physics-based models to Predictive Digital Twin via Interpretable Machine Learning},
pdfsubject={Digital Twin},
pdfauthor={Kapteyn, M.G., and Willcox, K.E.},
pdfkeywords={Digital twin, data-model fusion, model updating, reduced order model, unmanned aerial vehicle}
}
\titlespacing{\subsection}{0pt}{0.4\baselineskip}{0pt}
\title{From Physics-Based Models to Predictive Digital Twins via Interpretable Machine Learning}

\author{Michael G. Kapteyn\footnote{PhD Student, Department of Aeronautics and Astronautics, Student Member AIAA.}}
\affil{Massachusetts Institute of Technology, Cambridge, MA 02139}
\author{Karen E. Willcox \footnote{Director, Oden Institute for Computational Engineering and Sciences, Fellow AIAA}}
\affil{University of Texas at Austin, Austin, TX 78712}
\begin{document}
\maketitle
\begin{abstract}
\noindent This work develops a methodology for creating a data-driven digital twin from a library of physics-based models representing various asset states. The digital twin is updated using interpretable machine learning. Specifically, we use optimal trees---a recently developed scalable machine learning method---to train an interpretable data-driven classifier. Training data for the classifier are generated offline using simulated scenarios solved by the library of physics-based models. These data can be further augmented using experimental or other historical data. In operation, the classifier uses observational data from the asset to infer which physics-based models in the model library are the best candidates for the updated digital twin. The approach is demonstrated through the development of a structural digital twin for a 12ft wingspan unmanned aerial vehicle. This digital twin is built from a library of reduced-order models of the vehicle in a range of structural states. The data-driven digital twin dynamically updates in response to structural damage or degradation and enables the aircraft to replan a safe mission accordingly. Within this context, we study the performance of the optimal tree classifiers and demonstrate how their interpretability enables explainable structural assessments from sparse sensor measurements, and also informs optimal sensor placement.
\end{abstract}

\section{Nomenclature}
{\renewcommand\arraystretch{1.0}
\noindent\begin{longtable*}{@{}l @{\quad=\quad} l@{}}
$\alpha$ & complexity parameter of an optimal tree\\
$d_t$ & digital twin model at time $t$\\
$\boldsymbol{\epsilon}$ & vector of predicted strain measurements\\
$F$ & function mapping from models to observations\\
$\tilde{F}$ & noisy forward mapping from models to observations\\
$\mathbb{I}$ & identity matrix\\
$L$ & aerodynamic load factor\\
$\mu_1,\mu_2$ & model parameters governing the percentage reduction in stiffness in the damage regions\\
$M_j$ & label denoting of physics-based model $j$ from the model library\\
$\mathcal{M}$ & library of physics-based models\\
$\mathbf{M}$ & vector of models for each training datapoint \\
$\mathcal{N}$ & normal distribution\\
$n$ & number of training data points\\
$p$ & number of features (dimension of the observed data)\\
$p(\cdot|\cdot)$ & conditional probability density\\
$\mathbb{R}$ & the set of real numbers\\
$R(T)$ & misclassification error of the tree on the training data\\
$s$ & number of noise samples for each model\\
$T$ & an optimal classification tree that maps from observed data to models.\\
$|T|$ & depth of the classification tree\\
$\mathbf{v}^k$ & observation noise sample $k$\\
$\mathcal{V}$ & the space of all possible observation noise vectors\\
$\mathbf{x}_j$ & training vector of observed features from model $j$\\
$\mathbf{x}_j^k$ & training vector of noisy observed features from model $j$, with noise sample $k$\\
$\mathbf{\tilde{x}}_t$ & testing vector of observed features at time $t$\\
$\mathcal{X}$ & the space of all possible feature vectors\\
$\mathbf{X}$ & matrix containing all feature vectors in the training dataset
\end{longtable*}
Subscripts:
\noindent\begin{longtable*}{@{}l @{\quad=\quad} l@{}}
% $j$ & Model index \\
% $t$ & Timestep index \\
$j \in \{1,...,|\mathcal{M}|\}$ & model index \hspace{9cm} \\
$t \in \{0,1,2...\}$ & timestep index
\end{longtable*}
Superscripts:
\noindent\begin{longtable*}{@{}l @{\quad=\quad} l@{}}
% $k$ & Noise sample index\\
$k \in \{1,...,s\}$ & noise sample index\hspace{8.5cm}
\end{longtable*}}

\section{Introduction}\label{sec:intro}
This work develops an approach for creating predictive digital twins by leveraging interpretable machine learning methods to couple sensor data with physics-based models of the system. A library of physics-based models provides predictive capability---for any asset state represented in the library, we generate model-based predictions of observed quantities (e.g., quantities sensed onboard a vehicle). These predictions form a training set, to which we apply interpretable machine learning to train a classifier. Applying the classifier to online sensor data permits us to reliably infer which model best represents the current asset state, enabling us to maintain an up-to-date physics-based digital twin of a given asset.

Physics-based models offer a high degree of interpretability, reliability, and predictive capability and are commonplace throughout engineering. For example, during the development of an engineering system, such models are often created in order to explore the design space, and perform ``what-if'' analysis. These same models can also be used to analyze, predict, and optimize the performance of the system during operation. However, insights depend on the model used being an accurate reflection of the underlying physical system. The \textit{digital twin} paradigm aims to address this limitation by providing an adaptive, comprehensive, and authoritative digital model tailored to each unique physical asset. The digital twin paradigm has garnered attention in a range of engineering applications\cite{hartmann2020digital}, such as structural health monitoring and aircraft sustainment procedures \cite{glaessgen2012digital,li2017dynamic}, simulation-based vehicle certification \cite{glaessgen2012digital,tuegel2011reengineering}, and fleet management \cite{glaessgen2012digital,kraft2017engine,reifsnider2013multiphysics}. Some prior works have sought to create a digital twin by manually calibrating a system model to each physical asset at the beginning of the asset lifecycle\cite{jeon2019improving}. While these works serve to demonstrate the benefit of a digital twin, the manual calibration approach is limited to parameters that can be directly measured, or directly inferred based on a known relationship with measurable quantities. Manual calibration is also limited in scalability since the digital twin update is not automated.

Motivated by the proliferation of low-cost sensors and increasing connectivity between physical assets, in this work we leverage online sensor data to inform automatic and rapid dynamic adaptation of a physics-based digital twin, ensuring that it accurately reflects the evolving physical asset. We suppose one has a library of physics-based models, $\mathcal{M}$, where each model in the library represents a possible state of the physical asset. How this library of candidate models is constructed is outside of the scope of this paper, however, in this work we use as an example a library of component-based reduced-order models developed in prior work{\cite{kapteyn2020digitalttwin}. We show how such a library can be leveraged to generate a rich dataset of predictions, even for rare states or states that are yet to occur in practice such as failure modes of the asset. We then show how machine learning can be applied to this model-based dataset to train a data-driven model selector, $T$. This model selector maps observed data, $\mathbf{\tilde{x}}_t$, from the physical asset to estimate which model from the model library best explains the data, and should thus be used as the up-to-date digital twin, $d_t$, at time $t$. Thus, our approach to representing the digital twin is defined by the following equation:
\begin{equation}
d_t = T(\mathbf{\tilde{x}}_t) \in \mathcal{M}.
\label{eqn:master}
\end{equation}
Using this approach we have, at any time $t$, a reliable, predictive physics-based digital twin $d_t$, that is consistent with the most recent set of observations, $\mathbf{\tilde{x}}_t$, from the physical asset. In this work we address the development of the data-driven model selector, $T$, in a way that is tailored to the digital twin context.

In particular, ensuring that the digital twin is reliable demands that the data-driven model selector be both accurate and interpretable. Prior works on data-driven model updating have typically focused on  Bayesian methods\cite{kennedy2001bayesian,li2017dynamic,zhang2018aircraft,dourado2020bayesian,kapteyn2020digitalttwin}, or optimization-based machine learning methods, targeting either model fitting\cite{zakrajsek2017development}, model calibration\cite{zhao2019component}, or learning a model correction term\cite{chinesta2018virtual,yucesan2020hybrid}. While these methods can be accurate, they are typically treated as a black-boxes, meaning the digital twin model updating process is not interpretable or understandable, and offers little insight as to how the data are being used. To this end, we propose using a scalable approach for \textit{interpretable} machine learning based on optimal decision trees \cite{OCT,bertsimas2019machine}. In addition to achieving state-of-the-art prediction accuracy, this approach provides predictions that are interpretable because, via the partitioning of feature space, it is clear which observations or features are contributing to a prediction and where the decision boundaries lie. Furthermore, the optimal trees framework has a number of additional benefits in the digital twin context. The method naturally provides insight into which observations are most useful for a given prediction task. This benefit allows us to leverage the methodology for sensor placement, sensor scheduling, and risk-based inspection in support of the digital twin. The method also naturally extends to enable uncertainty quantification and risk-aware asset state estimation.

We demonstrate our methodology and illustrate the benefits of our contributions by means of a case study. We create a digital twin of a fixed-wing unmanned aerial vehicle (UAV). Our goal is for this digital twin to enable the UAV to become self-aware, in the sense that it is able to dynamically detect and adapt to changes in its structural health due to either damage or degradation \cite{allaire2012dynamic,lecerf2015methodology,singh2017methodology}.In this example, we build the digital twin from a previously developed library of component-based reduced-order structural models of the aircraft\cite{kapteyn2020digitalttwin}. Offline, we use this model library to create a dataset consisting of predicted sensor measurements for different UAV structural states. We use this dataset to train an optimal classification tree that identifies which sensor measurements are informative, and determines how these measurements should be used to determine the structural state of the UAV. Online, the UAV uses this classifier to rapidly adapt the digital twin based on acquired sensor data. The updated digital twin can then be used to decide whether to perform faster, more aggressive maneuvers, or fall back to more conservative maneuvers to minimize further damage or degradation.

The remainder of this paper is organized as follows. Section \ref{sec:optimaltrees} formulates the problem of data-driven model updating using a library of physics-based models and the optimal trees approach for interpretable machine learning. Section \ref{sec:OCTDT} discusses how optimal trees enable efficient and interpretable predictive digital twins. Section \ref{sec:results} presents the self-aware UAV case study which serves to demonstrate our approach. Finally, Section \ref{sec:conclusion} concludes the paper.

\section{Data-Driven Digital Twins via Interpretable Machine Learning}\label{sec:optimaltrees}
This section describes how interpretable machine learning is used in combination with a library of physics-based models to create predictive data-driven digital twins. Section \ref{sec:interpretableML} formulates the problem of data-driven digital twin model adaptation using a model library and machine learning. Section \ref{sec:OCT} presents an overview of interpretable machine learning via decision trees. Section \ref{sec:OCTconstruction} describes a recently developed approach for computing \textit{optimal decision trees}, and how we leverage this approach in the digital twin context.
\subsection{Problem formulation: Data-driven model selection}\label{sec:interpretableML}
We consider the challenge of solving \eqref{eqn:master} (see Sec.\ \ref{sec:intro}), which requires using observational data from a physical asset in order to determine which physics-based model is the best candidate for the digital twin of the asset, $d_t$, at time $t$. In particular, we suppose that during the operational phase of an asset we have access to $p$ sources of observational data that provide (often incomplete) knowledge about the underlying state of the asset. In general, these observations could be real valued (e.g., sensor readings, inspection data), or categorical (e.g., a fault detection system reporting \textit{nominal}, \textit{warning}, or \textit{faulty}, represented by integers $0, 1, 2$ respectively). We combine these observations into a so-called \textit{feature vector}, denoted by $\mathbf{\tilde{x}}_t\in\mathcal{X}$, where $t$ denotes a time index and $\mathcal{X}$ denotes the feature space, i.e., the space of all possible observations. We leverage these data to estimate which physics-based model best matches the physical asset, in the sense that it best explains the observed data. In this work, the set of candidate models we consider for the digital twin is denoted by $\mathcal{M}$. Thus, the task of data-driven model selection can be framed as an inverse problem, where we aim to find the inverse mapping from observed features to models, which we denote by
\begin{equation}
T: \mathcal{X} \rightarrow \mathcal{M}.
\label{eqn:Tmap}
\end{equation}
We derive this mapping using machine learning to define a model selector, $T$.

Training the model selector requires training data. To generate these training data, each model $M_j\in\mathcal{M}$ can be evaluated to predict the data, denoted by $\mathbf{x}_j$, that we would observe if the physical asset was perfectly represented by model $M_j$. This allows us to sample from the forward mapping
\begin{equation}
F: \mathcal{M} \rightarrow \mathcal{X},
\end{equation}
to generate $(\mathbf{x}_j, M_j)$ pairs for $j=1,...,|\mathcal{M}|$. It is often the case in practice that even if the asset were perfectly modeled by $M_j$, the data we actually observe are prone to corruption, e.g., due to sensor noise or inspection errors. We model this corruption as a random additive noise term drawn from a known noise model. An example of this would be additive Gaussian noise for a sensor with known bias and covariance. It is beneficial to account for this when training the model selector so that it can leverage this information, e.g., to learn that a given observation is typically too noisy to be reliably informative. We thus define the noisy forward mapping
\begin{equation}
\tilde{F}: \mathcal{M} \times \mathcal{V} \rightarrow \mathcal{X},
\label{eqn:noisymap}
\end{equation}
where $\mathcal{V}$ is the space of possible noise values. To sample from this noisy forward mapping, we first sample from the noise-free forward mapping to generate a pair $(\mathbf{x}_j, M_j)$, and then sample a noise vector $\mathbf{v}^k\in\mathcal{V}$ to generate a noisy prediction of the observed quantities, namely $\mathbf{x}_j^k = \mathbf{x}_j+\mathbf{v}^k$. Here the superscript $k$ denotes the index of the noise sample. We denote by $s$ the number of noisy samples we draw for each model $M_j$. The value of $s$ depends on the allowable size of the training dataset, which in turn depends on the computational resources available for training the machine learning model.

We sample training datapoints from the noisy forward mapping $\tilde{F}$. The resulting dataset takes the form $(\mathbf{x}_j^k, M_j)$, $j=1,...,|\mathcal{M}|$, $k=1,...,s$. As a final step, we vertically stack all of the training datapoints into a matrix representation $(\mathbf{X},\mathbf{M})$, where $\mathbf{X}\in\mathbb{R}^{n\times p}$ and $\mathbf{M}\in\mathbb{R}^n$. Here $n = s|\mathcal{M}|$ is the total number of datapoints in the training set. We use this dataset to learn the inverse mapping, and thus the model selector $T$.
\subsection{Interpretable machine learning using decision trees}\label{sec:OCT}
We now present an interpretable machine learning approach based on the notion of decision trees, and show how the training data described in the previous section can be used to train a classification tree that functions as the model selector, $T$, and enables accurate and interpretable data-driven digital twin model selection.

Decision trees are widely used in machine learning and statistics. Decision tree methods learn from the training data a recursive, hierarchical partitioning of the feature space, $\mathcal{X}$, into a set of disjoint subspaces. In this work we focus on classification trees, in which each region is assigned an outcome: either a deterministic or probabilistic label assignment. Outcomes are assigned based on the training data points in each region, i.e., the most common label in the deterministic case, or the proportion of points with a given label in the probabilistic case. To illustrate this idea, we use a simple demonstrative dataset\footnote{The data are taken from a classical dataset for classification problems: Fisher's iris dataset \cite{fisher1936use}, which is available from the UCI machine learning repository \cite{Dua:2019}}. In this example the training data $(\mathbf{X},\mathbf{M})$, consists of $n=150$ observations, where each $\mathbf{x}_j, j=1,...,n$ contains $p=4$ features, and there are three possible class labels $\mathcal{M} = \{M_1, M_2, M_3\}$.

In Figure \ref{fig:irispartition}, we plot this dataset according to the first two features, $x_1$ and $x_2$. Recall that in our context, each of these features would be a piece of observational data, for example a sensor measurement or the result of an inspection. We then show examples of using either axis-aligned or hyperplane splits to partition the feature space. We also show the decision trees corresponding to each partitioning.
\begin{figure}[h!]
\centering
\includegraphics[width=0.966\textwidth]{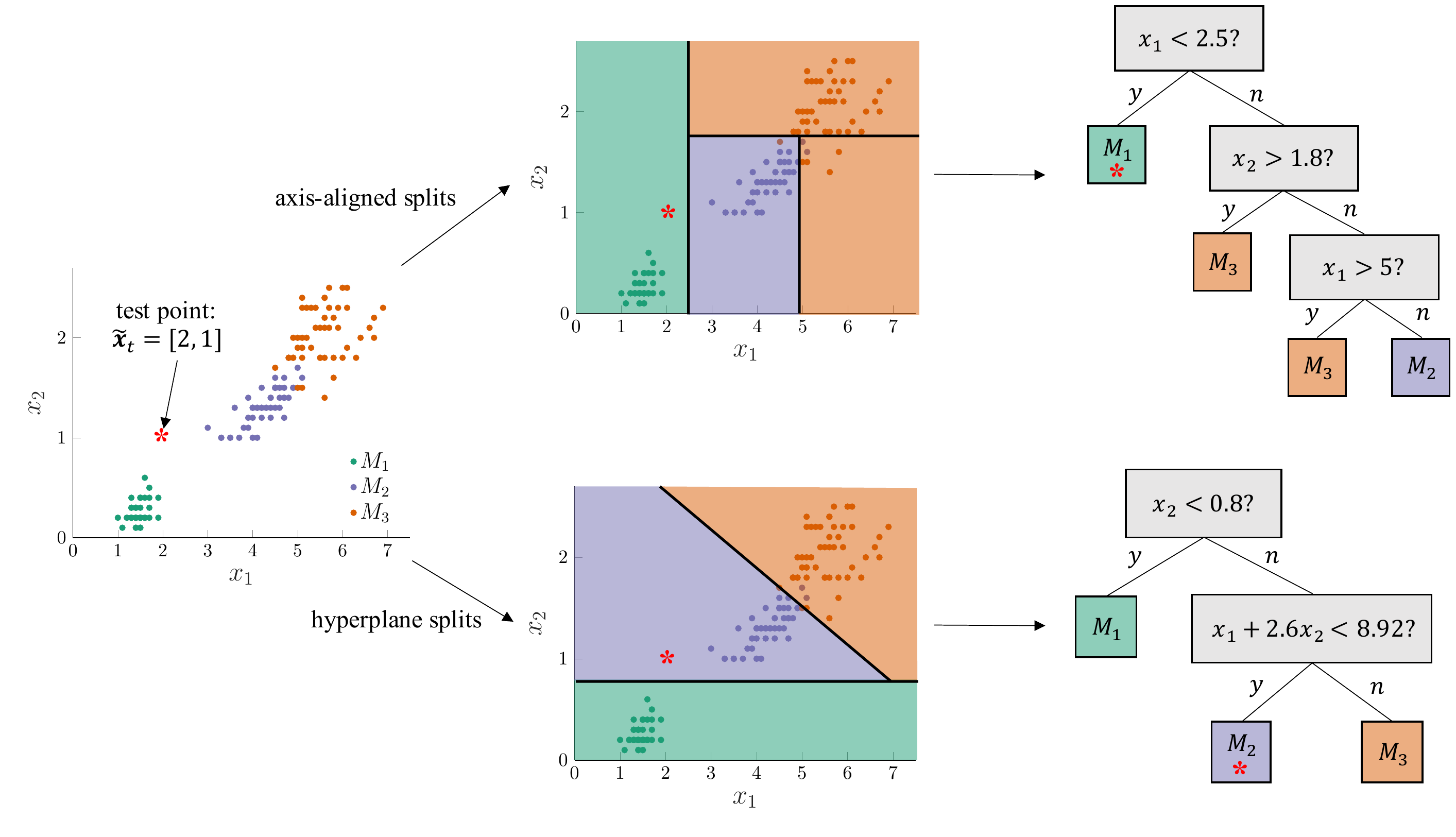}
\caption{Two possible ways to recursively partition the illustrative dataset, one using axis-aligned splits only, and another allowing for more general hyperplane splits. The corresponding classification tree is shown for each partitioning. A test point is shown to illustrate how a new datapoint is assigned a label using the classification tree.}
\label{fig:irispartition}
\end{figure}
Once a classification tree has been generated it can be used to predict outcomes for new, previously unseen feature vectors. When a test observation, $\mathbf{\tilde{x}}_t$, is obtained at some timestep $t$, evaluation begins at the root node of the tree. At each branching node in the tree, a decision is made depending on the observed feature values, until the test observation reaches a leaf node, at which point a predicted output is assigned. For example, the test observation $\mathbf{\tilde{x}}_t = [2, 1]$, depicted as an asterisk in Fig.\ \ref{fig:irispartition}, would be classified into $M_1$ using the axis-aligned splits and into $M_2$ using the hyperplane splits. Recall that in our context these labels would correspond to physics-based models, and the assigned label is the model that best explains the observation $\mathbf{\tilde{x}}_t$.

Hyperplane splits involve a linear combination of multiple features, and thus can be less interpretable than axis-aligned splits, but hyperplane splits have other advantages. In particular, since they are a more powerful modeling tool, trees with hyperplane splits can be shallower than trees with axis-aligned splits while achieving the same level of accuracy. A shallow tree with just a few hyperplane splits may be more interpretable than a deep tree with many axis-aligned splits, especially if the features used in the hyperplanes are somehow intuitively related. Therefore, incorporating hyperplane splits might in fact improve interpretability of the overall tree, while also improving accuracy for a given tree depth.
\subsection{Constructing an optimal decision tree}\label{sec:OCTconstruction}
We now present an overview of a recently developed scalable method for learning a so-called \textit{optimal classification tree (OCT)} from data, based on a formulation of the problem as a mixed-integer optimization problem\cite{OCT,bertsimas2019machine}. In contrast with previous methods such as CART\cite{CART} that construct sub-optimal trees using a greedy heuristic, the optimization approach simultaneously optimizes all splits in the tree, thereby avoiding the so-called lookahead pathology\cite{murthy1995lookahead}. Furthermore the optimization formulation naturally incorporates hyperplane splits, which have previously been incorporated using sub-optimal perturbation methods\cite{CART,SADT, OC1}. These improvements enable the optimal decision trees to achieving state-of-the-art accuracy comparable to ensemble tree methods such as Random Forests \cite{rf} or gradient boosted trees \cite{xgboost}, while maintaining the interpretability that comes from using only a single decision tree.

We follow the formulation given in \cite{OCT}, which involves framing the construction of an optimal decision tree as a mixed-integer optimization problem. The optimization variables include all the quantities necessary to define the tree, which we denote by $T$. This includes the coefficients defining the hyperplane splits for each decision node, as well as variables that determine the arrangement of nodes in the tree. The objective of the optimization problem is to construct a tree that minimizes the misclassification error on the training set, while also minimizing the complexity of the tree in order to improve interpretability and avoid overfitting, thereby promoting good out-of-sample performance. This objective function can be stated as:
\begin{equation}
\min\limits_{T} R(T) + \alpha|T|
\label{eqn:treeobj}
\end{equation}
where $T$ is the classification tree, $R(T)$ is the misclassification error of the tree on the training data, and $|T|$ is the complexity of the tree, as measured by the number of splits. The complexity parameter $\alpha$ governs the tradeoff between accuracy and complexity of the tree. The full problem, including the definition of variables comprising the tree, $T$, and the constraints required to enforce the tree structure can be found in \cite{OCT}. This optimization problem can solved directly using commercial solvers such as Gurobi \cite{gurobi}, using solutions from heuristic methods such as CART as warm starts\cite{OCT}.

The mixed-integer optimization approach has the benefit that it can produce solutions that are certifiably globally optimal. However, this approach generally scales inefficiently, as the number of integer decision variables increases exponentially as the depth of the tree increases, and linearly as the number of training data points increases. To overcome this, an efficient local-search method has been developed\cite{bertsimas2019machine} which is able to efficiently solve the problem to near global optimality for practical problem sizes, in times comparable to previous methods. In particular, under a set of realistic assumptions, the cost of finding an OCT using the local-search is only a factor of $\log(n)$ greater than CART. In addition to being optimal in-sample, it is shown in \cite{OCT} that OCT outperforms CART and Random Forests in terms of out-of-sample accuracy, across a range of $53$ practical classification problems from the UCI machine learning repository \cite{Dua:2019}, and at all practical values for the tree depth.

Moreover, the mixed-integer programming formulation and local-search procedure both naturally extend efficiently to incorporate hyperplane splits, producing a so-called \textit{optimal classification tree with hyperplane splits (OCT-H)}. The formulation is flexible in that it accommodates constraints on the split complexity, in particular the number of variables, or which specific combinations of variables are allowed in each hyperplane split, as well as constraints that the coefficients on each variable in the hyperplane be integer; this can help to ensure that the splits remain interpretable. Note that in the limit of allowing a split complexity of only one variable per split OCT-H reduces to the OCT case. The incorporation of hyperplane splits does not significantly increase the computational cost of the algorithm, but improves the out-of-sample accuracy to state-of-the-art levels. In particular, across the $53$ datasets from the UCI repository, OCT-H was able to significantly outperform CART, Random Forests, and OCT, with significantly shallower (less complex) trees \cite{OCT}. OCT-H also outperforms gradient boosted trees at depths above three \cite{bertsimas2019machine}. As a final comparison to other state-of-the-art methods, we note that \cite{bertsimas2019machine} compares optimal trees with various neural network architectures. In particular, they show that various types of feedforward, convolutional, and recurrent neural networks can be reformulated exactly as classification trees with hyperplane splits. This result suggests that the modeling power of an OCT-H is at least as strong as these types of neural network.

\section{Optimal trees as an enabler of digital twins}\label{sec:OCTDT}
This section highlights the desirable characteristics of optimal trees, and argues that an optimal classification tree (OCT or OCT-H) is an ideal candidate for the data-driven model selector, $T$, introduced in \eqref{eqn:master}.
\subsection{Efficient and interpretable digital twin model updating}
In this work we train an optimal decision tree, $T$, using the training data, $(\mathbf{X},\mathbf{M})$, generated using the set of digital twin candidate models, $\mathcal{M}$, as described in Section \ref{sec:interpretableML}. The result is a classification tree, $T$, that fulfills the function of the data-driven model selector introduced in Eqn.\ \eqref{eqn:master} and defined in Eqn.\ \eqref{eqn:Tmap}. In particular, when a noisy measurement is observed from the physical asset it is passed as input to the optimal classification tree. The optimal classification tree then uses the structure it has learned from the physics-based model in order to provide a prediction for the model, $M_j\in\mathcal{M}$, which best represents the physical asset based on the observed measurement.

An important feature of our approach is that we generate training data using the physics-based model library. Using physics-based models to simulate sensor responses to various asset states in this way allows one to avoid expensive and time-consuming experiments. It also means that we can include in the training data asset states that are previously unseen in the real-world, such as anomalous or rare events, for which experimental data are limited or unavailable. Of course, if historical data are available, they can easily be added to the training dataset as additional samples. Note that generating the training data in this way is a many-query task, as it requires evaluating every model in the model library. This is a potentially expensive offline computation, but can be accelerated using reduced-order modeling techniques\cite{kapteyn2020digitalttwin} so that each model is fast to evaluate. Once an optimal tree has been trained in the offline phase, performing a classification online is extremely fast. This is important in applications in which dynamic decision-making requires rapid digital twin updates in response to online data streams (see for example, the case study given in Sec.\ \ref{sec:results}).

Another important feature is that predictions made via an optimal tree are interpretable, in the sense that it is clear which features in the observed data are contributing to a decision and what the decision boundaries are on these features. In the digital twin context, this means the optimal tree reveals exactly which observations are being used and exactly how they are being used in order to update the digital twin. In contrast with black-box approaches this type of interpretable classifier makes it possible to diagnose anomalous or unexpected decisions and thus makes the predictions more trustworthy and reliable. It also helps practitioners to understand the rationale behind each decision, even if they do not understand how the decision tree itself was generated.
\subsection{Informing optimal sparse sensing strategies}\label{sec:sensorplacement}
The optimal trees framework naturally incorporates optimal feature or sensor selection, since solving for the optimal tree automatically reveals which of the observed features are most informative for a given classification task. In particular, while the training dataset contains $p$ different sources of observational data, the resulting decision trees typically utilize only a subset of these features. For example, Figure \ref{fig:irispartition} shows that for this illustrative dataset only two out of four features (namely, $x_1$ and $x_2$) are required for an accurate decision tree with axis-aligned splits.

Note that situations in which features are observed simultaneously require that all features appearing in the decision tree be observed at the outset. On the other hand, if features are acquired or processed sequentially we can achieve even greater sparsity by only measuring features if they are required to make the relevant classification, i.e., only those appearing in a single root-to-leaf path through the tree. For example, suppose that acquiring each of the features in our illustrative dataset (Figure \ref{fig:irispartition}) required performing an expensive manual inspection of a physical asset. If we were to use the classification tree with axis-aligned splits to update the assets digital twin, we would begin by performing a single inspection and measuring only feature $x_1$. If $x_1< 2.5$, then we would classify the digital twin state as $M_1$, and feature $x_2$ would not be required. If, on the other hand, $x_1 > 2.5$, we would then proceed to a second inspection that measures feature $x_2$, before classifying the digital twin as either $M_2$ or $M_3$. Note however, that the optimal tree we show here is not the only optimal solution. In fact, the split given by $x_1< 2.5$ could be moved to the end of the sequence; this corresponds to a different, but still optimal partition. A potential benefit of this solution is that the classification label $M_3$ is moved further up in the tree, and could only require one feature ($x_2$). This could be favorable, e.g. if $M_3$ represented a critical asset state that required rapid detection, whereas $M_1$ was less critical, or if feature $x_2$ was much cheaper to acquire than $x_1$. In practice we could compare different optimal solutions in this way as a post-processing step, or we could modify the optimization problem to reward the appearance of critical labels higher up in the tree. This example serves to highlight how the optimal decision trees can be used to reveal accurate sparse sensing strategies, and can prioritize critical and/or cost effective classifications.

We have thus far assumed that the set of $p$ available observations is fixed, for example due to fixed sensor locations. In many situations the set of observations is not fixed a priori. This is the case if we are in the process of designing a new digital-twin-enabled asset, or retrofitting an existing physical asset with new sensors or inspection schedules to enable a digital twin. In such cases, the optimal trees methodology provides a valuable tool for informing optimal sensor placement. In particular, instead of using a fixed set of pre-existing sensors, we could instead specify a (potentially large) set of possible observations, for example candidate sensor types and sensor locations. We would then proceed in the usual way by using the physics-based models to generate the training dataset by simulating how each sensor would respond to different asset states. From this dataset we train an optimal tree which reveals the set of candidate observations which would be most informative for estimating the asset state. The observations which appear in the resulting decision tree would then be selected as the sensors to be installed. This functionality is further illustrated in Section \ref{sec:fixedcandidate}. Note that this approach naturally extends to allow multi-modal sensors; one simply needs the ability to simulate the response of each sensor to various asset states. Sensor-related cost functions or constraints on the sensor budget could also be incorporated into the sensor placement problem in order to trade off cost versus estimation performance for different sensor architectures.
\subsection{Incorporating uncertainty quantification and risk}
Thus far the output of the optimal tree model selector has been treated as a deterministic estimate of the single model that best reflects the input feature vector. This is limited in that it does not account for any uncertainty in the estimate. Here we show how this limitation can be addressed by modifying the tree so that it outputs a distribution over the models in the model library, rather than deterministically estimating a single model.

During the training phase, part of our objective is to minimize the misclassification error of the tree on the training data. Another way to view this is that the optimization seeks a tree that maximizes the training label purity in each leaf, i.e., it aims to find a tree structure that classifies training datapoints such that each leaf contains datapoints with similar labels. In reality, due to partial observability and/or sensor noise, it is often impossible to find a perfect partitioning of the training dataset for any reasonable tree complexity. In the deterministic case, the algorithm simply selects the most common training label in each leaf, and assigns this to be the output for any datapoint classified into that leaf. For example, consider a situation in which 75\% of the training datapoints classified into a given leaf of the tree have label $M_1$, while the remaining 25\% have label $M_2$. In the deterministic case, when a test point, $\tilde{\mathbf{x}}_t$ is classified into this leaf, the tree would assign it label $M_1$. This can be modified so that the tree instead outputs a distribution over the possible labels, namely
\begin{equation}
p(d_t = M_j | \mathbf{\tilde{x}}_t) = \begin{cases}
     0.75 & \text{for }j=1\\
     0.25 & \text{for }j=2
    \end{cases}
\end{equation}
In this way the decision tree not only estimates the best candidate model but also quantifies the uncertainty in this estimate, and thus the resulting uncertainty in the digital twin. High uncertainty in the digital twin could indicate that the model library is inadequate for representing the state of the physical asset. Thus this could be used to trigger further investigation and subsequent model library enrichment. The quantification of uncertainty in this way also enables the optimal tree to be incorporated into a wider Bayesian framework, wherein one could incorporate prior beliefs about the asset state and/or observations.

Furthermore, one could also incorporate a measure of risk into the optimal tree, for example, the probability of failure for each asset state. This is motivated by the fact that often different assets states are not treated equally, so one might prioritize the ability of the digital twin to detect critical states rather than minimizing the digital twin uncertainty overall. For example, it might be more important that the digital twin is accurate when an asset is nearing a failure or other decision boundary. We can account for this in the optimization problem used to generate the decision tree by computing the misclassification error using a weighted sum over training samples. Weighting samples by their associated risk or probability of failure has the effect of prioritizing the ability of the tree to accurately estimate high-risk states, while placing relatively less emphasis on safe states. In this way we optimize the digital twin's ability to accurately quantify the probability of failure of the asset.

\section{Case Study: Toward a Self-Aware UAV}\label{sec:results}
This section presents a case study that serves to demonstrate the approach described in the previous two sections. Section \ref{sec:UAVhardware} presents a fixed-wing UAV that serves as the subject of this case study and describes the library of physics-based models for the UAV structure which serves as the basis for its digital twin. In Section \ref{sec:UAVprocedure} we outline how this model library enables us to generate a rich set of training data comprised of predicted sensor measurements for different structural states represented in the model library. It also describes how we use this training data to generate optimal trees that enable data-driven adaptation of the digital twin, and how we evaluate the performance of the trees. Section \ref{sec:UAVadapt} then presents the optimal trees generated for this case-study and explores their performance. Finally, Section \ref{sec:UAVsim} presents simulation results demonstrating how the optimal trees could be used in a rapidly updating digital twin that enables the UAV to respond intelligently to damage or degradation detected in the wing structure.
\subsection{Physical asset and physics-based model library}\label{sec:UAVhardware}
Our physical asset for this research is a 12-foot wingspan, fixed-wing UAV. The fuselage is from an off-the-shelf Telemaster airplane, while the wings are custom-built with plywood ribs and a carbon fiber skin. The electric motor and avionics are also a custom installation. The wings of the aircraft are designed to be interchangeable so that the aircraft can be flown with pristine wings or wings inflicted with varying degrees of damage. This allows us to collect in-flight structural sensor data for multiple structural states, and test whether a digital twin of the aircraft is capable of dynamically detecting changes in the state using structural sensor data. Photos of the aircraft during a recent series of flight tests are shown in Figure \ref{fig:aircraftphotos}.
\begin{figure}[h]
\begin{center}
\includegraphics[height=1.9in]{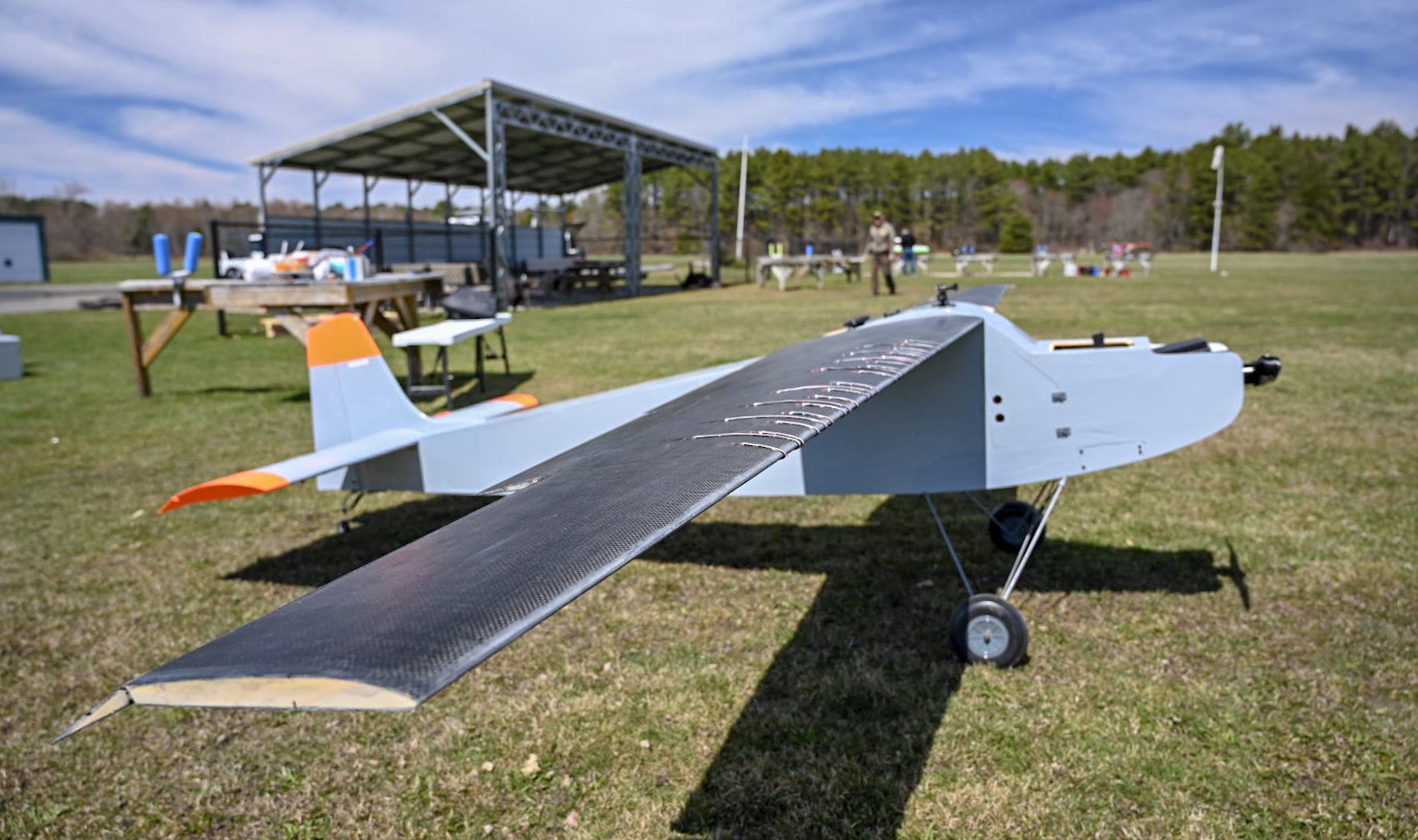}
\adjustbox{trim={.05\width} {0.1\height} {0.05\width} {.05\height},clip}%
{\includegraphics[height=2.235in]{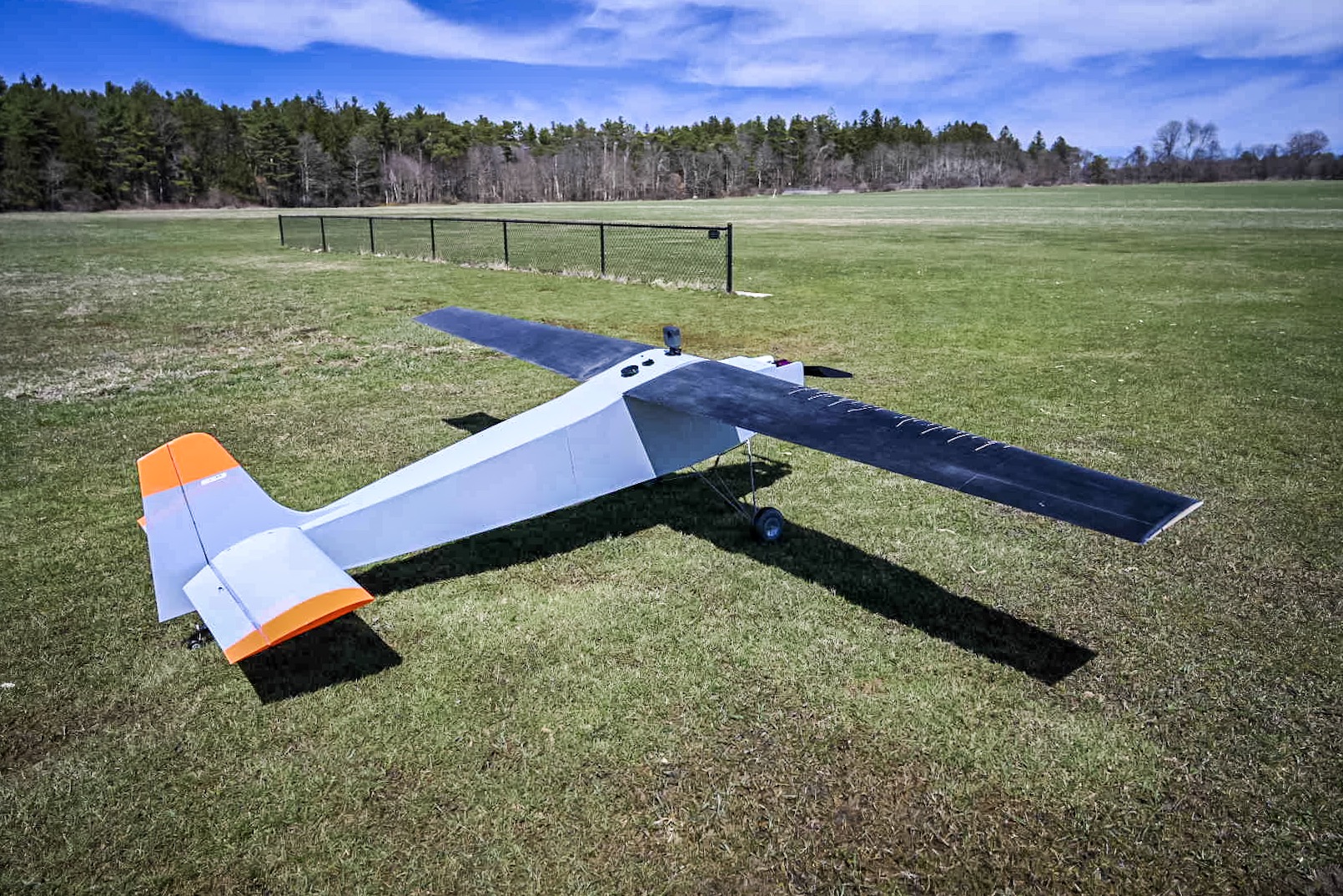}}
\end{center}
\caption{The custom-built hardware testbed used in this research. We create a digital twin of this 12-foot wingspan aircraft, and update the digital twin in response to online data from structural sensors on the aircraft wings.}
\label{fig:aircraftphotos}
\end{figure}
In this case study, we consider an illustrative scenario in which the right wing of the UAV has been subjected to structural damage or degradation in flight. This change in structural state could undermine the structural integrity of the wing and reduce the maximum allowable load, thereby shrinking the flight envelope of the aircraft. A self-aware UAV should be capable of detecting and estimating the extent of this structural change, so that it can update the flight envelope and replan its mission accordingly. In this case study we enable this self-awareness through a data-driven physics-based digital twin.

The physics-based model library that serves as the foundation of the UAV digital twin in this case-study is presented in prior work\cite{kapteyn2020digitalttwin}. We provide a high-level overview of the model library herein, and refer the reader to these works for further details. The model library is derived from a component-based reduced-order model for solving the linear elasticity equations on the full 3D UAV airframe. Figure \ref{fig:wingstructure} details the structure of the wing as represented in the model.
\begin{figure}[h]
\centering
\includegraphics[width=0.775\textwidth]{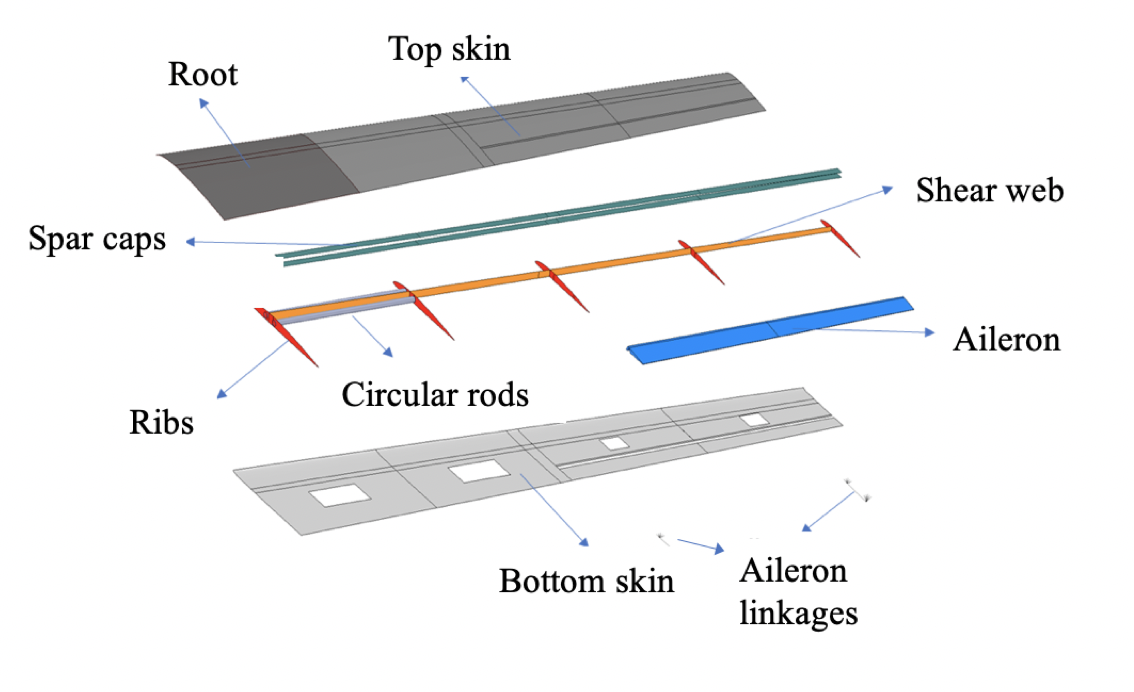}
\includegraphics[width=0.775\textwidth]{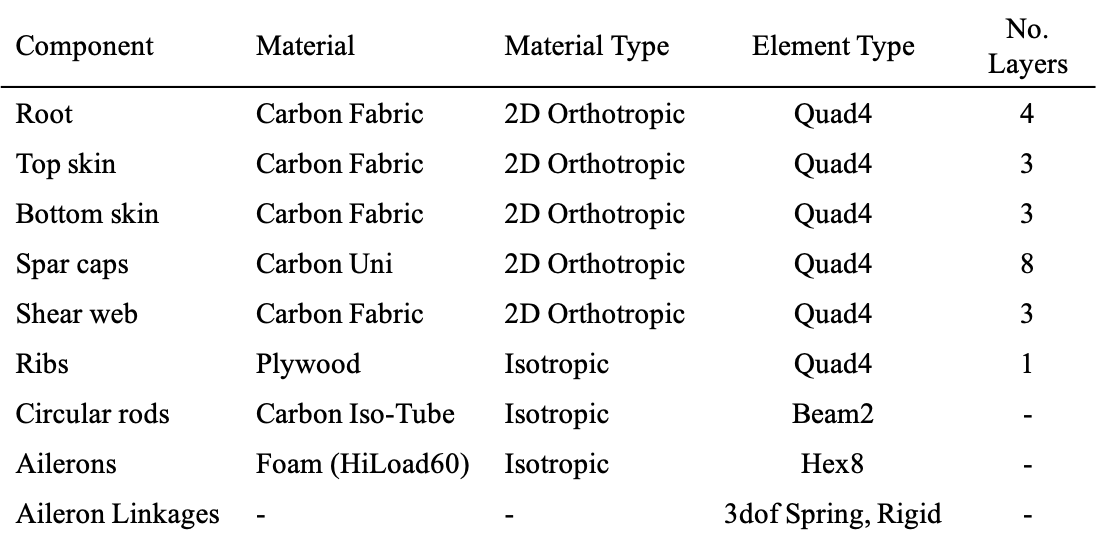}
\caption{The internal structure of the aircraft wing. We use a combination of material properties and element types in order to capture the level of detail required to accurately model structural health in our digital twin.}
\label{fig:wingstructure}
\end{figure}

We model a changed structural state (e.g., caused by damage, degradation, or some other structural event) through an effective reduction in stiffness. This model contains 28 so-called ``effective damage regions'' distributed across the airframe. For each effective damage region we introduce a scalar parameter governing the percentage reduction in material stiffness within the region. To illustrate the proposed methodology we will restrict our focus to just two of the effective damage regions on the aircraft, but note that the methodology would naturally scale to accommodate more damage regions. The two effective damage regions we focus on are located on the right-wing of the aircraft as shown in Figure \ref{fig:damagelibrary}.
\begin{figure}[h]
\begin{center}
\includegraphics[width=0.8\textwidth]{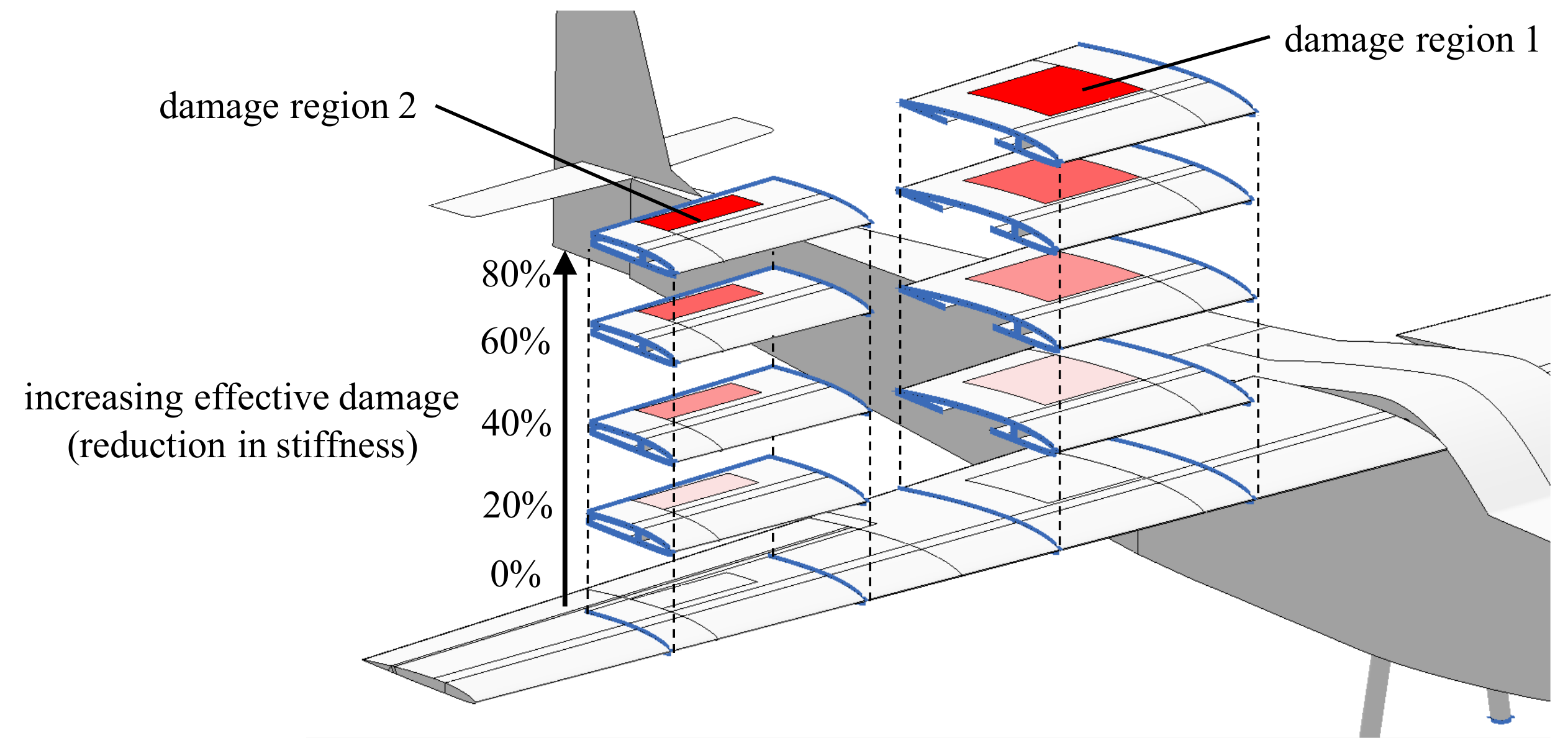}
\end{center}
\caption{An illustration of the model library used in this case study. We sample five values of the stiffness parameter in each of the effective damage regions (highlighted red). The model library is constructed by taking all combinations of stiffness parameters for the two components.}
\label{fig:damagelibrary}
\end{figure}
We denote the model parameters, namely the percentage stiffness reduction in damage regions 1 and 2, by $\mu_1$, and $\mu_2$ respectively. To construct the model library, $\mathcal{M}$, we first sample five linearly spaced values of each parameter, corresponding to a reduction in stiffness in the damage regions of between $0\%$ and $80\%$. We then take all combinations of the two model parameters to generate a model library, $\mathcal{M}$, containing $|\mathcal{M}|=25$ unique physics-based models, each representing a different UAV structural state. This shows that even for our restricted model library with only two damage regions, our digital twin of the UAV would be able to adapt to $25$ different structural states.

The data-driven digital twin is enabled by using in-flight sensor data to classify the current structural state of the UAV into one of the states represented in the model library $\mathcal{M}$. The sensors we consider are uniaxial strain gauges mounted on the top surface of the right wing. We consider as a baseline the sensor suite installed on the physical UAV asset, as shown in Figure \ref{fig:aircraftphotos}. In particular, we consider 24 uniaxial strain gauges distributed in two span-wise rows on either side of the main spar between 25\% and 75\% span. Figure \ref{fig:wingschematic} shows a schematic of the right wing with sensor locations. For this case-study, we exclude sensors 17, 18, 23 and 24 (shown as gray circles in Fig. \ref{fig:wingschematic}), since these sensors are located in the damage regions.
\begin{figure}[h]
\begin{center}
\includegraphics[width=0.9\textwidth]{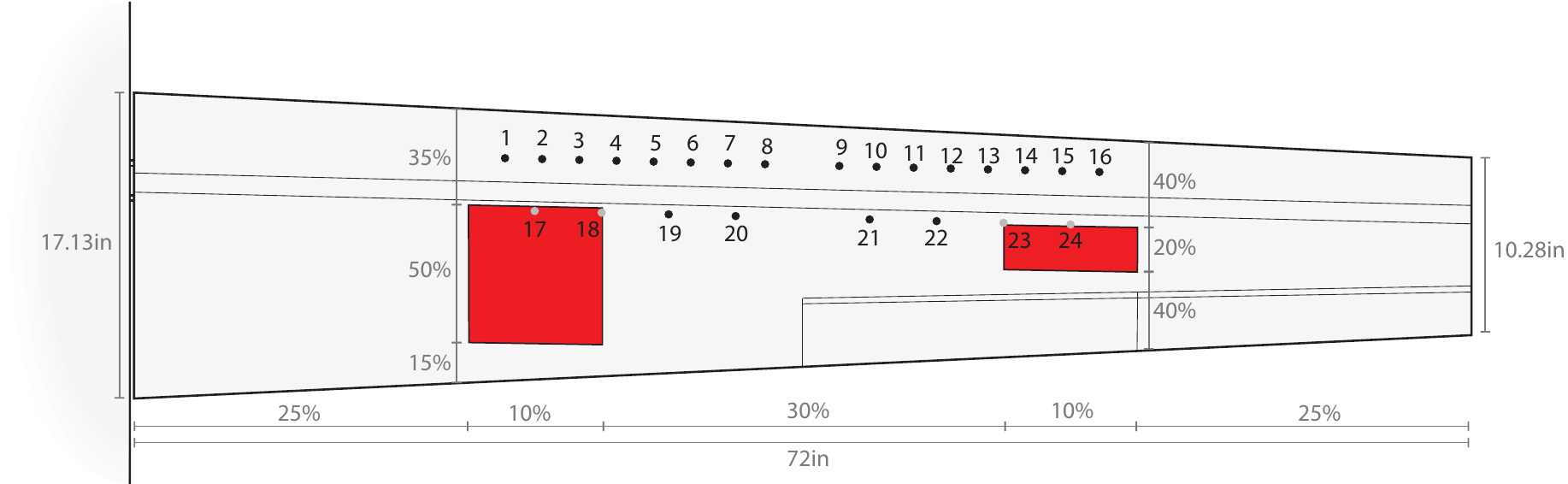}
\end{center}
\caption{Schematic of the UAV wing. The two damage regions we consider in this case-study are highlighted in red. Uniaxial strain gauges are shown as circles and are referred to by their integer label. Sensors 17, 18, 23 and 24 are excluded from this case-study.}
\label{fig:wingschematic}
\end{figure}
\subsection{Procedure for training and evaluation of optimal trees}\label{sec:UAVprocedure}
This describes the procedures we adopt for generating and evaluating optimal trees for the UAV digital twin case-study.
\subsubsection{Generation of training data}
Our goal is to leverage online structural data from the $20$ wing-mounted strain gauges to rapidly classify the state of the UAV into the model library, $\mathcal{M}$. In this scenario our observations are noisy strain measurements at strain gauge locations. We use units of microstrain throughout and, since the strains encountered are typically compressive, we treat compressive strains as positive for simplicity. These measurements take the form
\begin{equation}
\boldsymbol{\epsilon}(M_j,L) + \mathbf{v},
\end{equation}
where $\boldsymbol{\epsilon}$ is a vector representing the true strain at each of the $24$ strain gauge locations. The strain depends on the load factor, $L$, (ratio of lift force to aircraft weight), and the UAV structural state, which we describe by a model $M_j$. In this illustration we model the sensor noise as zero-mean white noise, $\mathbf{v}^k$, so that
\begin{equation}
\mathbf{v}^k \sim \mathcal{N}(\mathbf{0}, 1000\mathbb{I}),
\label{eqn:noise}
\end{equation}
where $\mathcal{N}$ denotes a Gaussian distribution and $\mathbb{I}\in\mathbb{R}^{20\times20}$ is an identity matrix.

Our structural model of the UAV is based on linear elasticity, so the strain predicted by the model is linear with respect to the applied load factor. Since the load factor is typically known based on the aircraft kinematics, we normalize measurements by the load factor and define the observed data to be a vector of load-normalized strains of form
\begin{equation}
\mathbf{x}_j^k = \frac{\boldsymbol{\epsilon}(M_j, L)+\mathbf{v}^k}{L}.
\label{eqn:strain}
\end{equation}
We generate training data of this form for every model in the model library, $M_j, j=1,...,|\mathcal{M}|$,  by drawing samples from the noisy forward mapping, $\tilde{F}$ (defined in Eqn.\ \ref{eqn:noisymap}). The procedure for drawing samples is as follows: We set $L=3$ (representing a $3g$ pull-up maneuver) and compute the corresponding aerodynamic load using an ASWING \cite{drela1999integrated} model of the UAV. We then apply this load to a structural reduced-order model $M_j$ and compute the strain that would be measured by each strain gauge. Finally, we draw a random noise sample $\mathbf{v}_k$, according to \eqref{eqn:noise}, and add this to the computed strain, before normalizing by the load factor to give a feature vector $\mathbf{x}_j^k$, consisting of $p=24$ noisy load-normalized strain measurements. The final step is repeated for $s=100$ noise samples $k=1,...,s$. Note that generating the training data in this way is a many-query task, requiring $|\mathcal{M}|$ evaluations of the UAV structural model. This is a one-time, offline procedure, and in our case is accelerated by the use of reduced-order models in the model library.

Our goal is to use load-normalized strain data in order to classify the UAV structural health into a structural state represented by one of the $|\mathcal{M}| = 25$ models in the library. Directly classifying the UAV structural health into one of the library models requires a tree of at least depth five, so that it has at least $25$ possible labels (one for each leaf in the tree). Since each model, $M_j$, has two parameters, $\mu_1$, and $\mu_2$, we choose to train a separate optimal tree classifier for each parameter. This is so that each tree requires only $5$ possible labels (a depth of at least three). This is simply to make the resulting trees more interpretable, since the choice of model $M_j$ can be easily inferred from the two parameters. In particular, note that we can easily recover a single tree for directly classifying $M_j$ by appending the tree for classifying $\mu_2$ to every leaf of the tree for classifying $\mu_1$. The resulting tree predicts both $\mu_1$ and $\mu_2$ for a given input, which corresponds uniquely to a model $M_j$. To this end we define the two outputs of interest to be the parameters defined in model $M_j$, which are denoted by $[\mu_1,\mu_2]_j$.

This process results in a dataset containing a total of $n=2500$ datapoints. Each datapoint is of the form $(\mathbf{x}_j^k,[\mu_1,\mu_2]_j)$. The feature vector, $\mathbf{x}_j^k$ is a prediction of the $p=24$ noisy load-normalized strain measurements corresponding to the structural state represented by the two parameters, $\mu_1$ and $\mu_2$, which can take one of $5$ discrete values corresponding to 0\%, 20\%, ..., 80\% reduction in stiffness respectively.
\subsubsection{Training an optimal tree classifier}
To generate optimal trees we use a Julia implementation of the local-search algorithm described in Sec.\ \ref{sec:OCTconstruction}, provided in the Interpretable AI software package \cite{InterpretableAI}. This implementation allows one to specify a number of parameters and constraints to be used in the optimization problem  (Eqn.\ \ref{eqn:treeobj}). We set the complexity trade-off parameter, $\alpha$, to zero and the minimum number of training points in each leaf of the tree to one. To control the complexity of the resulting tree we explicitly constrain the maximum depth and the maximum number of features appearing in each split, which we refer to as the split complexity. This approach allows us to explore the tradeoff between classification accuracy and complexity in a controlled way, by generating the tree with optimal classification accuracy for a given constrained complexity. Note that in practice these parameters do not need to be set explicitly, and can instead by found using standard cross-validation techniques.
\subsubsection{Evaluating the performance of an optimal tree classifier}
To evaluate the performance of an optimal tree classifier, we follow standard machine learning practice and hold out 30\% of the data ($750$ datapoints) as a testing set. We feed each test datapoint into the classifiers to determine its estimated structural state, given by the parameter $\mu_1$ or $\mu_2$, and compare this with the structural state that was used to generate that test point. We quantify the performance of each optimal tree by computing the mean absolute error (MAE) between the true percentage reduction in stiffness, and the reduction in stiffness estimated by the optimal tree classifier.
\subsection{Results: Data-driven model selection via optimal trees}\label{sec:UAVadapt}
In this section we apply the optimal trees methodology described in Section \ref{sec:optimaltrees} to the training data described in the previous section. We demonstrate how optimal trees serve as an interpretable classifier that enables a data-driven digital twin. First, Section \ref{sec:intuition} demonstrates how optimal trees are generated for our UAV application. Then, Section \ref{sec:depthsparsity} explores the trade-off between classifier interpretability and classification performance for optimal trees. Finally, Section \ref{sec:fixedcandidate} demonstrates how the optimal trees methodology could be used for optimal sensor placement, and how this can provide performance benefits over an ad-hoc approach.
\subsubsection{Illustrative results}\label{sec:intuition}
We begin by presenting illustrative results to demonstrate how the optimal tree classifiers in this case study correspond to a partitioning of the feature space, i.e. the space of all possible structural sensor measurements from the UAV. We also show how the structure of the tree corresponds to a sparse selection of sensors that are required to classify the structural state. To this end, we start by constraining the optimal trees to have a maximum depth of three and to utilize at-most two sensors. This is so that the resulting trees are more easily interpreted, and so that the corresponding partitioning of the feature space can be easily visualized on a two-dimensional subspace (each dimension corresponding to one of the two chosen sensors).

We first focus on classifying the first model parameter, $\mu_1$ using axis-aligned splits, i.e. we constrain the maximum split sparsity to one. The resulting OCT trained for this task utilizes sensors 1 and 2, and is shown in Figure \ref{fig:OCT1}. Also shown is the corresponding partitioning of the feature space and the locations of these sensors in relation to the damage regions.
\begin{figure}[h]
\begin{center}
\includegraphics[align=c, trim={0 4.5cm 1cm 5cm},clip, width=0.5\textwidth]{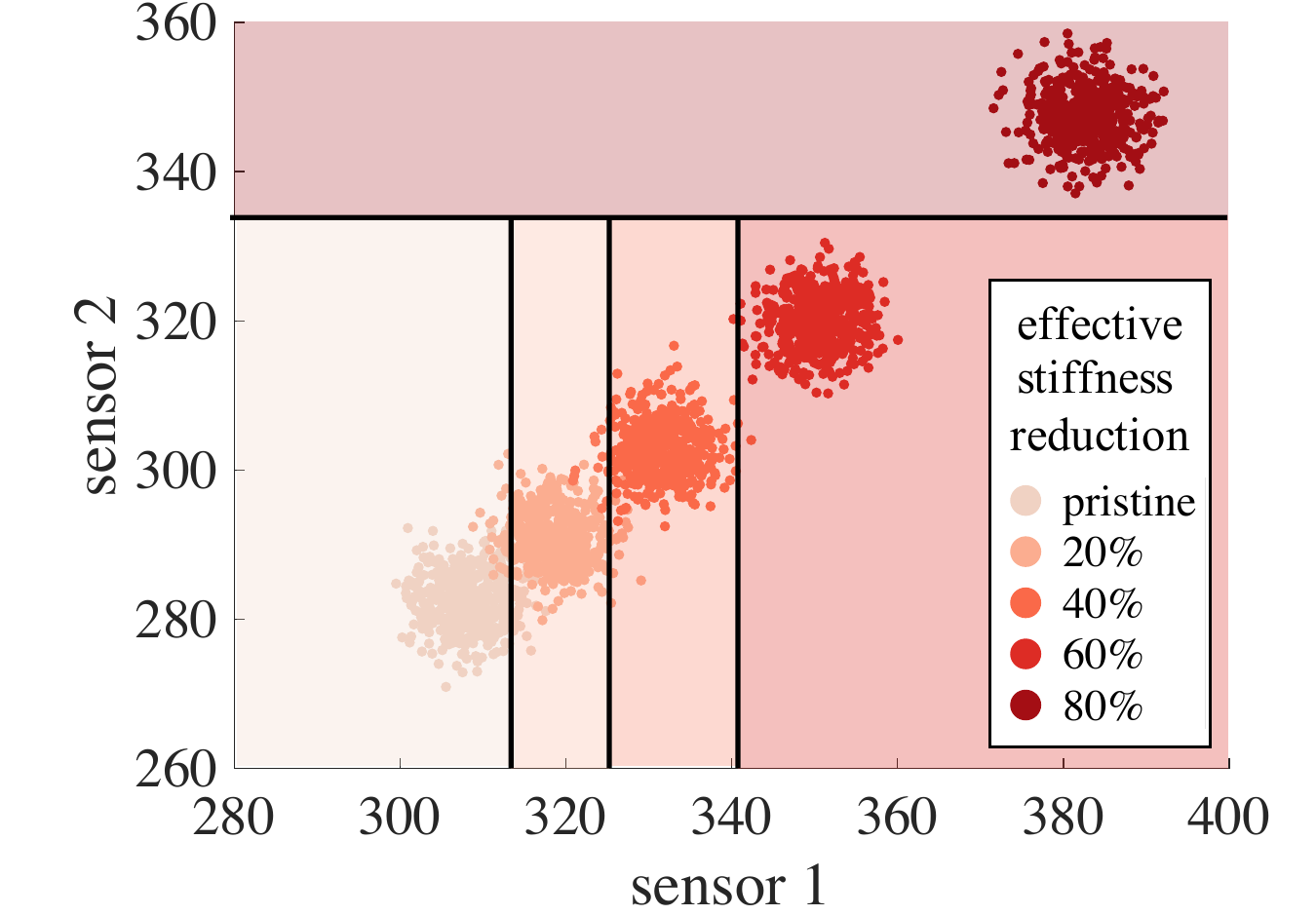}
\includegraphics[align=c, width=0.45\textwidth]{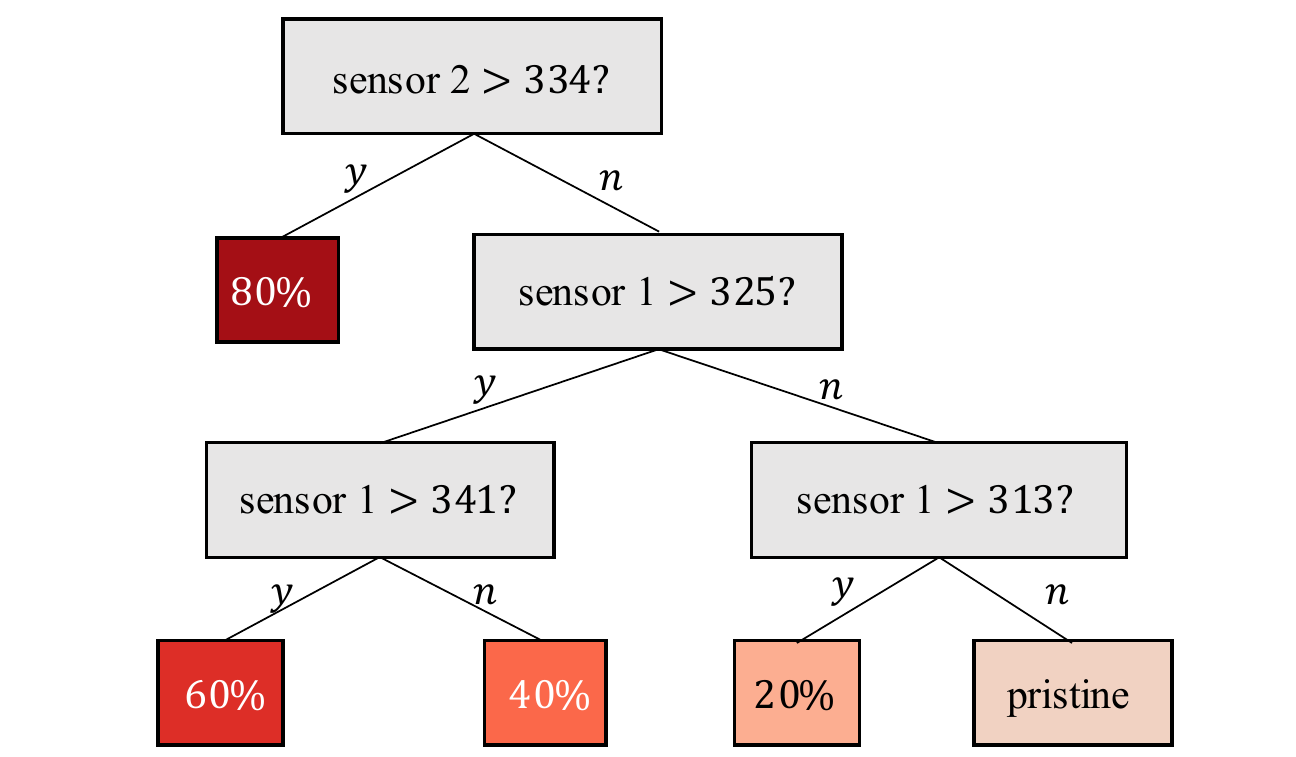}
\includegraphics[width=0.9\textwidth]{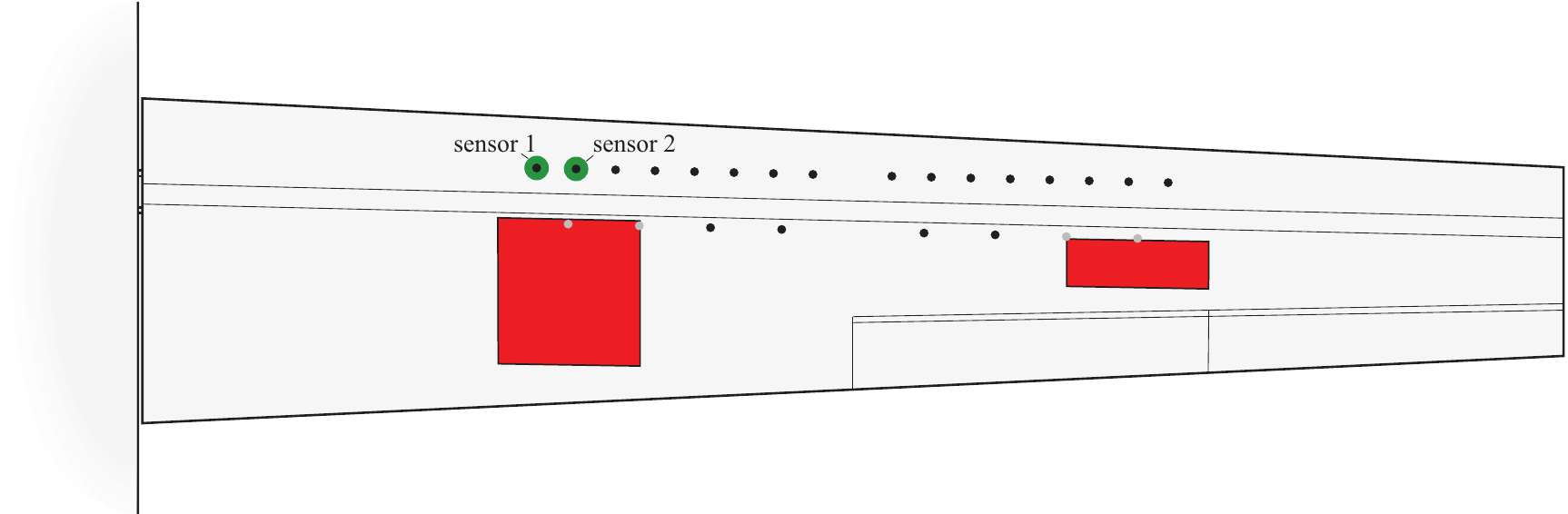}
\end{center}
\caption{An OCT for computing the model parameter $\mu_1$. Left: Partitioning of the feature space (the space of strain measurements) using axis-aligned splits. Right: The decision tree for classifying the value of $\mu_1$.}
\label{fig:OCT1}
\end{figure}
The MAE for this classifier is $0.65$ on the training set, which corresponds to $57$ out of the $1750$ training datapoints being misclassified (a misclassification rate of $3.3\%$). The MAE is only slightly higher for the test set at $0.88$. This shows that just sensors 1 and 2 are sufficient for accurately classifying the reduction in stiffness in damage region 1. This can be explained physically by the fact that these two sensors are close to the damage region. Note however, that no explicit information about proximity is present in the training data. Instead, the optimization problem for generating the optimal tree finds that sensors 1 and 2 are simply the most informative for this classification decision. We see in Figure \ref{fig:OCT1} that the predicted strain increases with the reduction in stiffness parameter equally for both sensor 1 and 2. This suggests that axis-aligned splits might not be suited to this problem. Figure \ref{fig:OCTH1} shows the results obtained when the maximum split sparsity is increased to two, i.e. we now allow hyperplane splits involving sensors 1 and 2.
\begin{figure}[H]
\begin{center}
\includegraphics[align=c, trim={0 4.5cm 1cm 5cm},clip, width=0.5\textwidth]{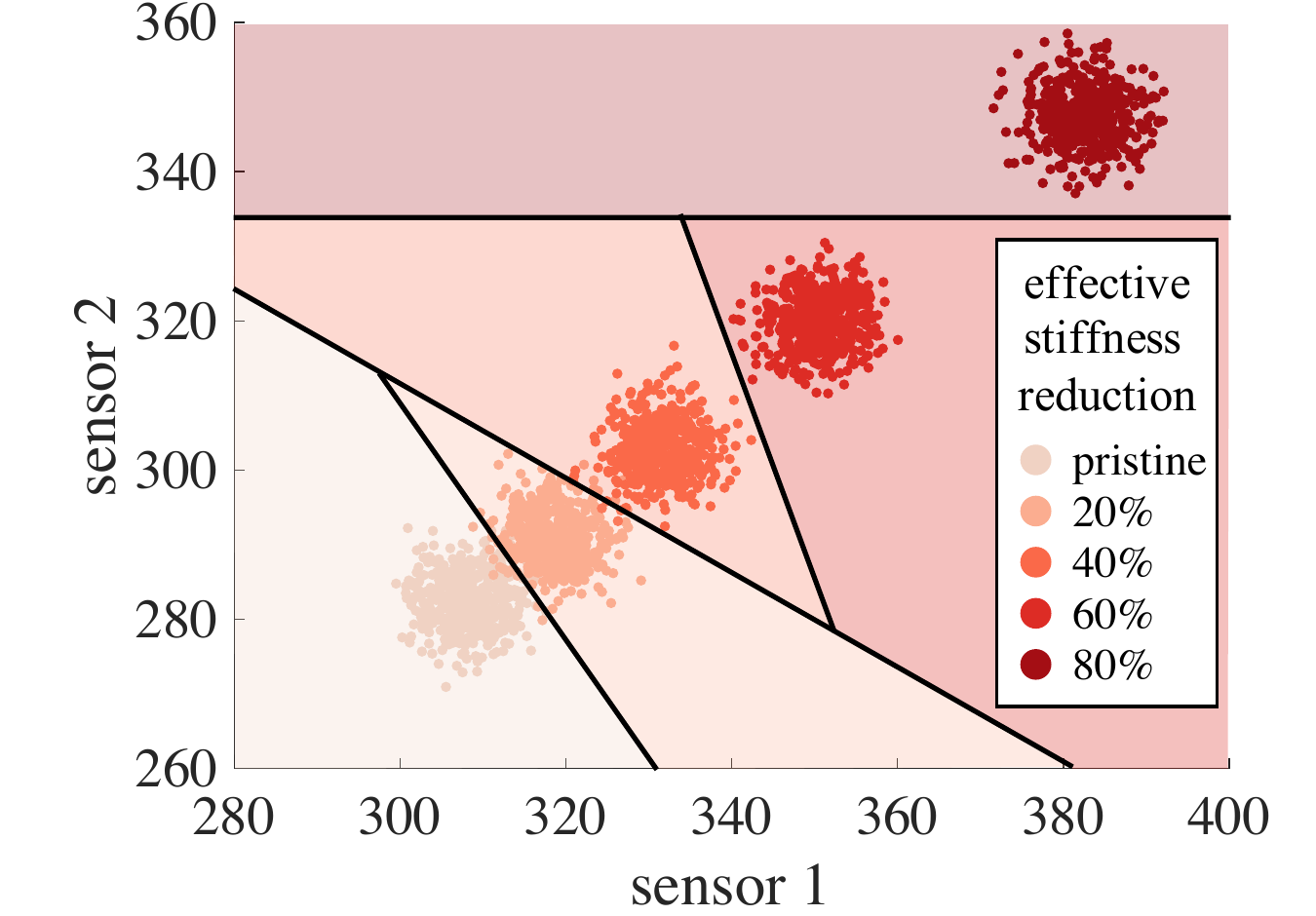}
\includegraphics[align=c, width=0.45\textwidth]{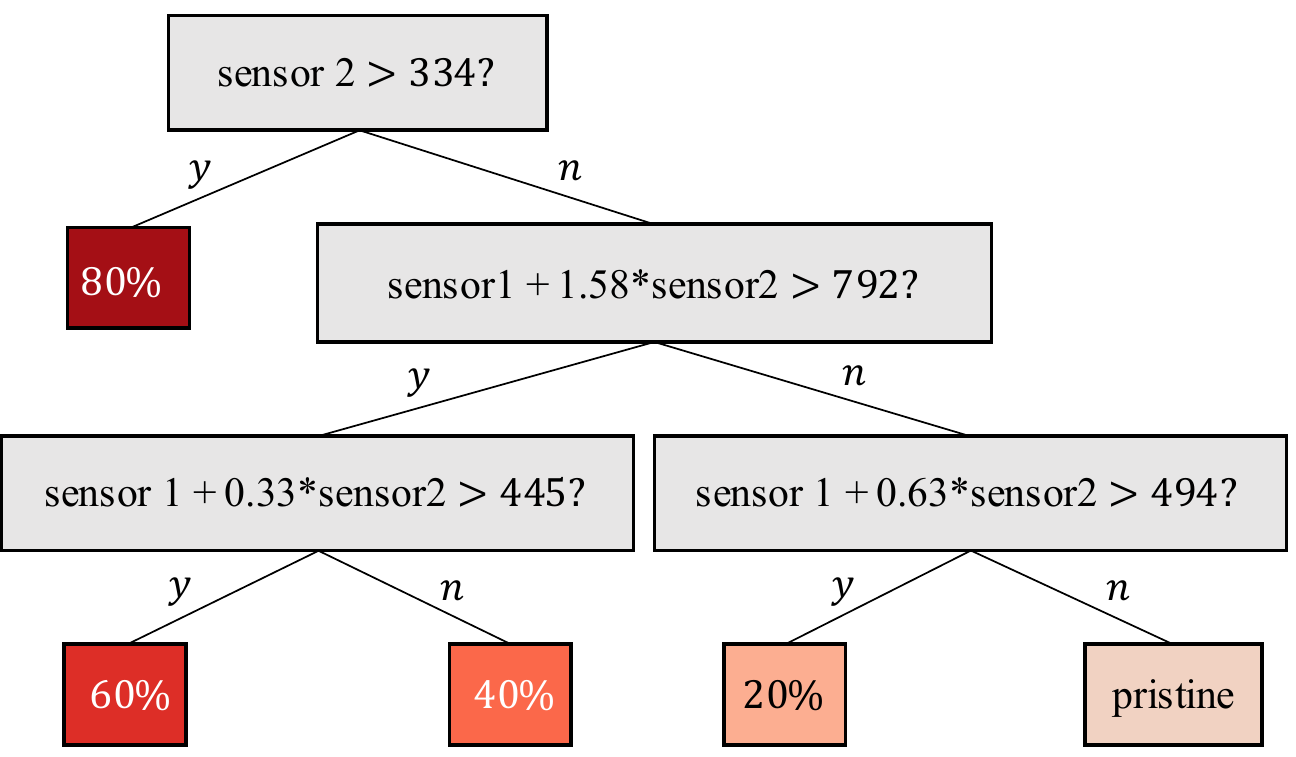}
\end{center}
\caption{An OCT-H for computing the model parameter $\mu_1$. Left: Partitioning of the feature space (the space of strain measurements) using axis-aligned splits. Right: The decision tree for classifying the value of $\mu_1$.}
\label{fig:OCTH1}
\end{figure}
In this case the MAE improves to $0.22$ on the training set, which corresponds to just $20$ out of the $1750$ training datapoints being misclassified (a misclassification rate of $1.1\%$). The MAE on the test set also improves to $0.4$. This highlights how increasing the maximum split sparsity can improve the classification performance of the tree, at the cost of making the classifier more complex, and thus less interpretable. This trade-off is further explored in Sec.\ \ref{sec:depthsparsity}.

Classifying the second model parameter, $\mu_2$, is more challenging since the second damage region is smaller than the first, and is nearer to the tip of the wing, thus having less of an effect on the wing deflection and the resulting strain field. The OCT trained for this task is shown in Figure \ref{fig:OCT2}, along with the partitioning of the feature space and the location of the utilized sensors in relation to damage region 2.
\begin{figure}[H]
\begin{center}
\includegraphics[align=c, trim={0 4.5cm 1cm 5cm},clip, width=0.5\textwidth]{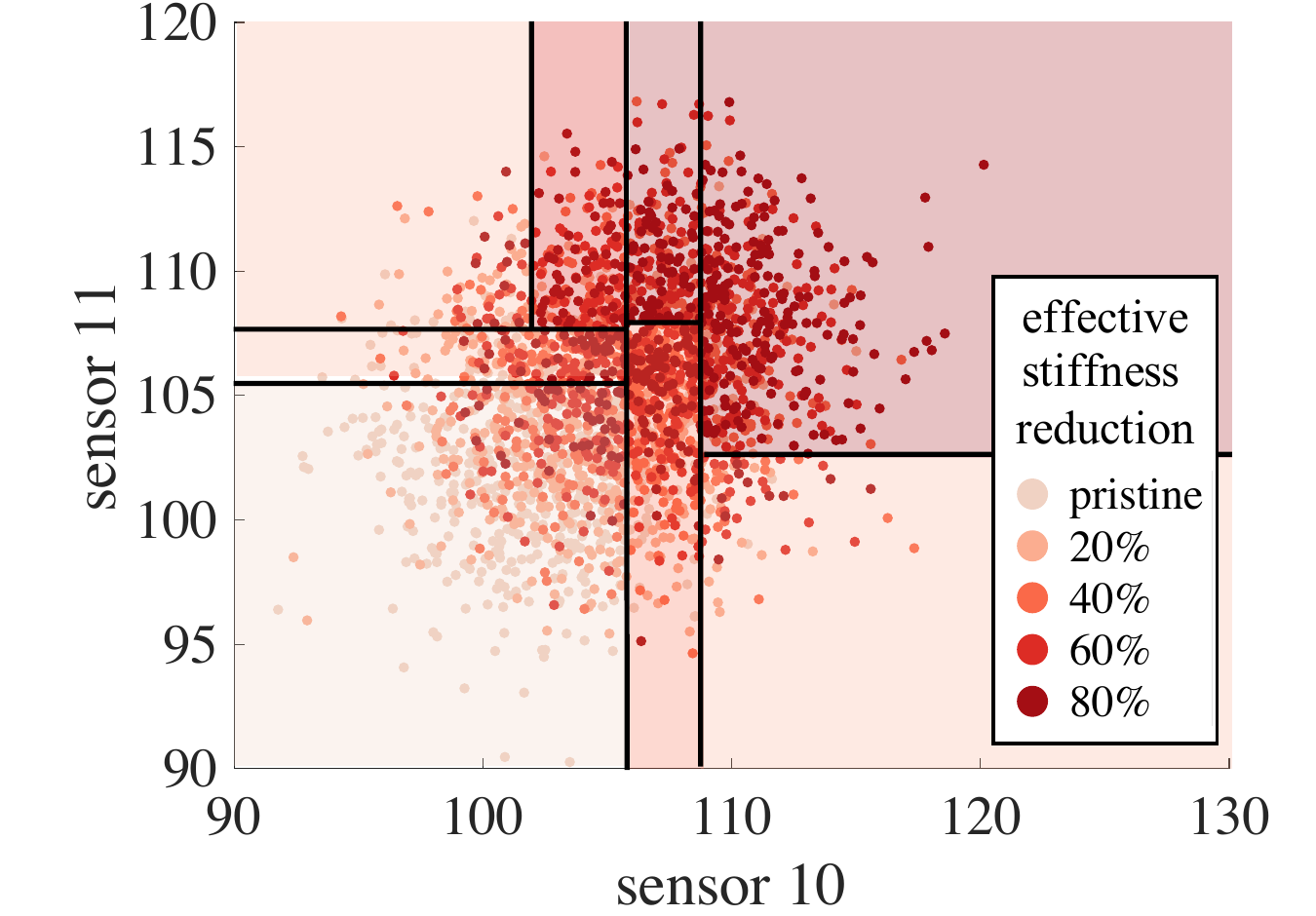}
\includegraphics[align=c, width=0.45\textwidth]{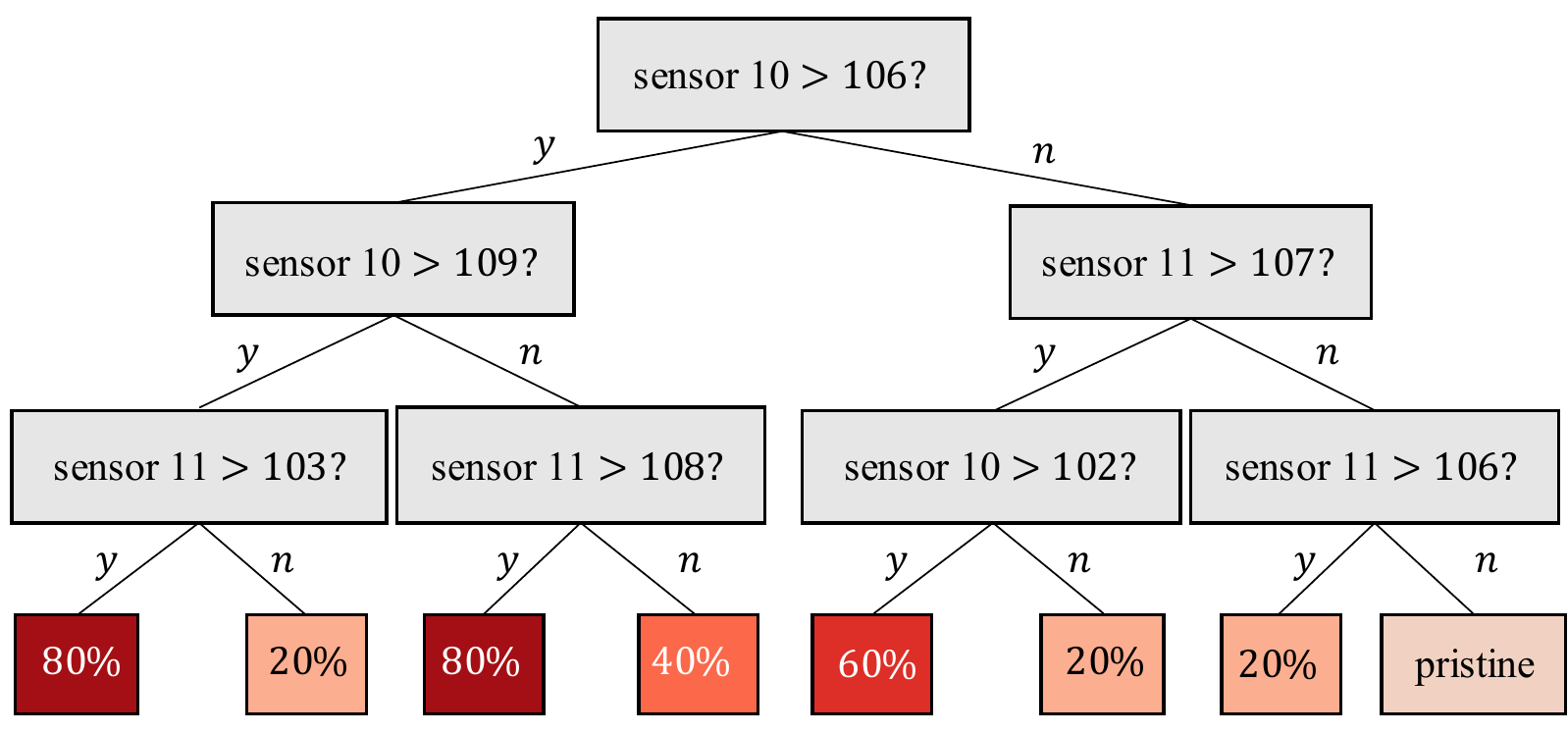}
\includegraphics[width=0.9\textwidth]{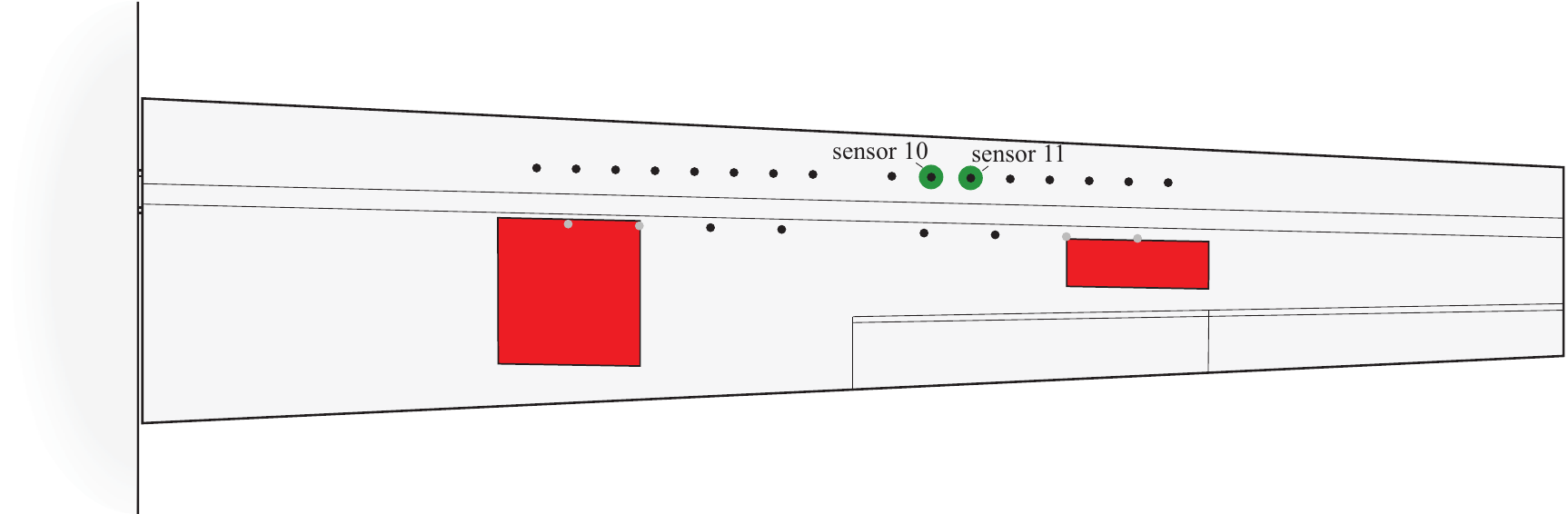}
\end{center}
\caption{An OCT for computing the model parameter $\mu_2$. Left: Partitioning of the feature space (the space of strain measurements) using axis-aligned splits. Right: The decision tree for classifying the value of $\mu_2$.}
\label{fig:OCT2}
\end{figure}
In this case, the MAE is 19.44 on the training set and 20.7 on the test set. This corresponds to a high misclassification rate: 60.1\% and 64.4\% respectively. This high error is due to the fact that the datapoints corresponding to different parameter values are not easily separable using just two sensors (as seen in Fig. \ref{fig:OCT2}). Improving this error would require either a more complex tree, or improving our sensing capability. We explore the latter option further in Sec.\ \ref{sec:fixedcandidate}. %
\subsubsection{Classifier performance vs. complexity}\label{sec:depthsparsity}
In this section we explore the tradeoff between classification performance and complexity of the optimal trees used in the UAV digital twin. Figure \ref{fig:sec02depthsparsity} shows the MAE on the training and test sets for optimal trees trained to classify $\mu_1$ with varying tree depth and split complexities.
\begin{figure}[h]
\begin{center}
\includegraphics[width=0.9\textwidth]{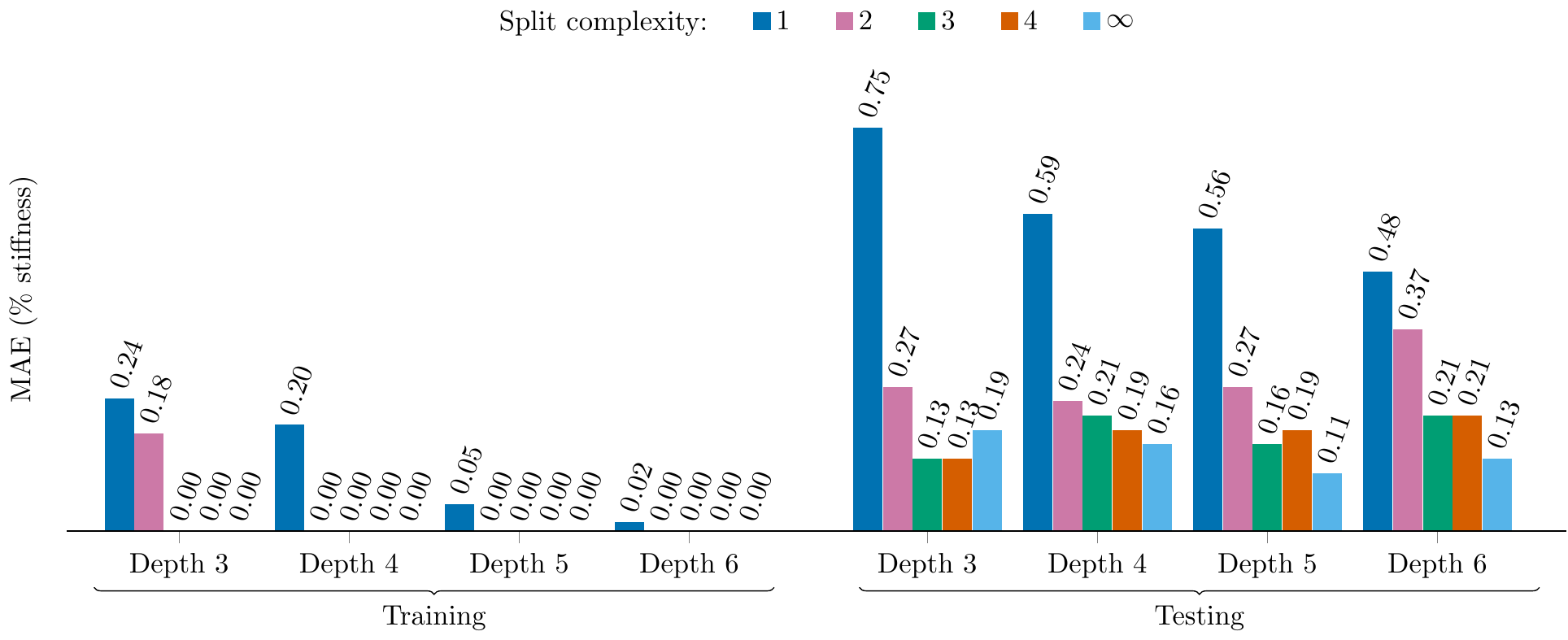}
\end{center}
\caption{Plot of classification performance (MAE) versus tree depth and split complexity for optimal trees trained to classify $\mu_1$, the model parameter governing the reduction in stiffness in damage region 1}
\label{fig:sec02depthsparsity}
\end{figure}
We see that with a tree depth of at least four and a split complexity of at least two the optimal tree classifier is able to fit the training data exactly. Performance on the test set is marginally lower in all cases, but is still very strong.

Figure \ref{fig:sec04depthsparsity} repeats this analysis, this time for optimal trees trained for the more difficult task of classifying the reduction in stiffness in damage region 2. Note the difference in axis scaling when compared with Figure \ref{fig:sec02depthsparsity}.
\begin{figure}[H]
\begin{center}
\includegraphics[width=0.9\textwidth]{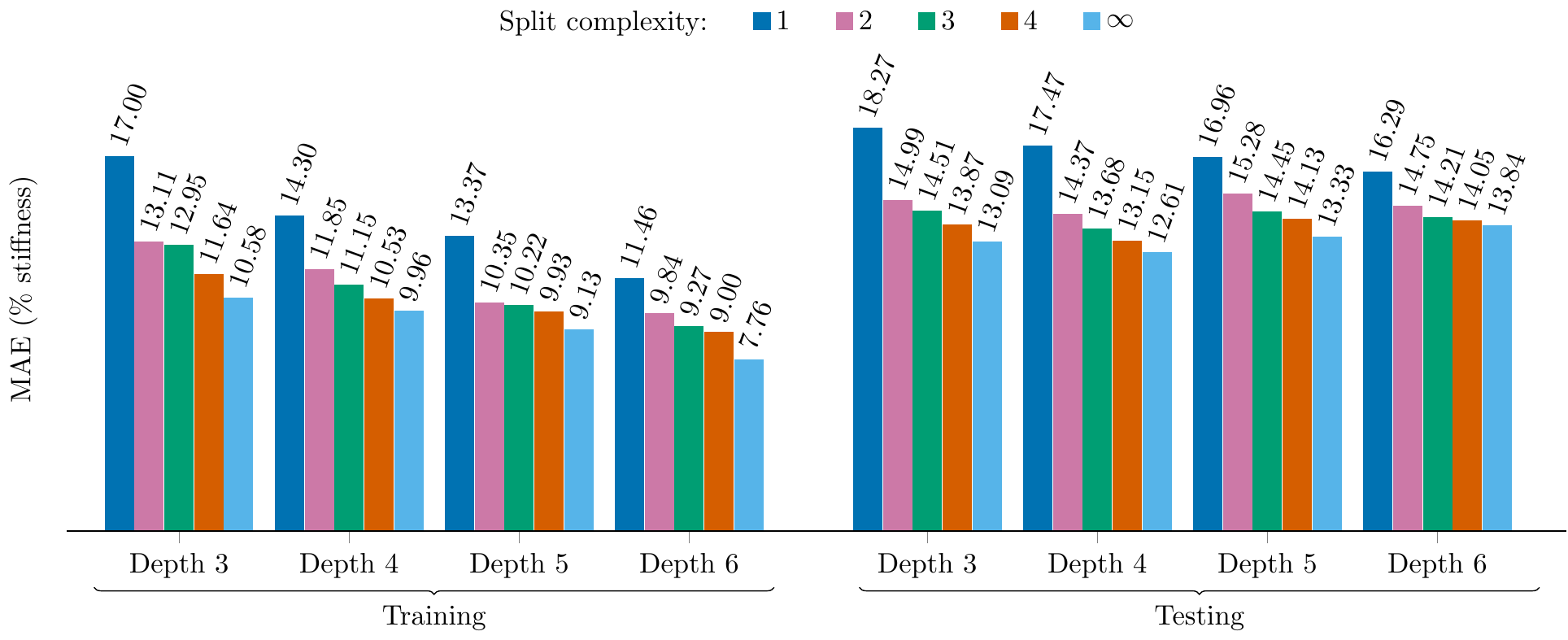}
\end{center}
\caption{Plot of classification performance (MAE) versus tree depth and split complexity for optimal trees trained to classify $\mu_2$, the model parameter governing the reduction in stiffness in damage region 2}
\label{fig:sec04depthsparsity}
\end{figure}
Here we see that the ability of the optimal tree to fit the training data increases with both depth and sparsity, achieving an MAE of $7.76$ with a tree of depth six and unlimited split complexity - a reduction in error of roughly 54\% compared with the depth three tree using axis-aligned splits. The out-of-sample performance on the testing set improves uniformly with split complexity, but does not improve uniformly with depth. We see that the out-of-sample performance is best at depth four, suggesting that the depth five and six trees are over-fitting the training data.
\subsubsection{Improving classification performance via optimal sensor placement}\label{sec:fixedcandidate}
In this section we demonstrate how optimal trees can scale to a large number of features and how this enables the methodology to be used for sensor placement or sensor selection, as described in Section \ref{sec:sensorplacement}.

In the previous analysis, we utilized the sensor suite which is currently installed on the hardware testbed. The strain gauges installed on the aircraft are not positioned optimally for the classification task we consider in this case-study. Instead, the sensor placement was determined through an ad-hoc analysis whereby the pristine wing finite element model was solved, and regions of high strain were identified. Note that the optimal sensor placements would be areas where the strain field is most sensitive changes in the structural state. This is not necessarily the same as the regions where the strain magnitude is highest.

A more principled approach to positioning the sensors involves leveraging optimal trees to directly determine which sensors are most informative for updating the digital twin. We do this by first determining a large set of feasible sensor locations and then training optimal classification trees using these candidate sensors as features. To illustrate this, we consider a set of 58 candidate sensors in addition to the 24 sensors already installed. The candidate sensors are arranged in four spanwise rows along the wing, as shown in Figure \ref{fig:candidatesensors}. As with the fixed sensor locations, we exclude sensor locations that occur within a damage region, giving a total of 67 possible sensor locations, and thus 67 features to inform classification in the optimal tree. The sensors highlighted in green are those that are utilized by the depth three tree trained using the candidate sensors, which is described further in the results that follow.
\begin{figure}[h]
\begin{center}
\includegraphics[width=0.9\textwidth]{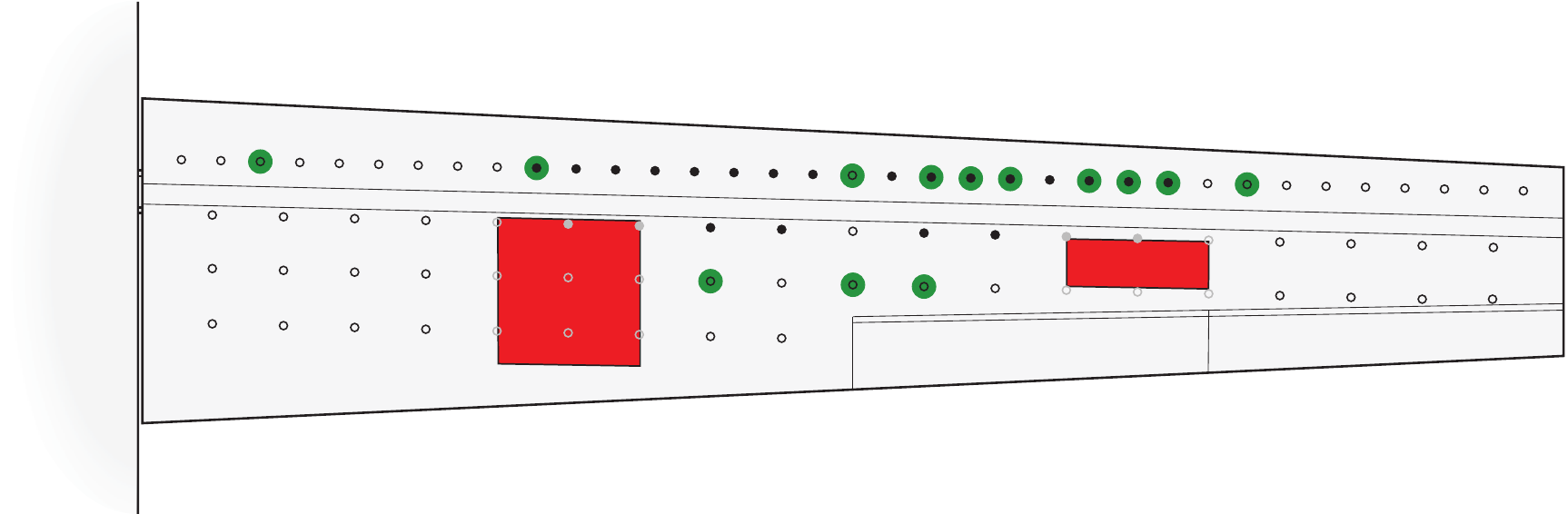}
\end{center}
\caption{Schematic of the UAV wing showing candidate sensor locations. The two damage regions we consider in this case-study are highlighted in red. Currently installed strain gauge sensors are shown as filled circles, while candidate sensor locations are shown as open circles.}
\label{fig:candidatesensors}
\end{figure}
We adapt the methodology described in Section \ref{sec:UAVprocedure} to generate a dataset that includes 67 features corresponding to the complete set of candidate sensors. We then train optimal trees to classify $\mu_2$. Figure \ref{fig:sec04fixedcandidate} shows the MAE for optimal trees of varying depth trained using the candidate sensor locations compared with using the already installed sensors. To focus the comparison on the positioning of sensors rather than the quantity of available sensors, we limit the total number of sensors appearing in each tree by setting the hyperplane complexity to four.
\begin{figure}[h]
\begin{center}
\includegraphics[width=0.8\textwidth]{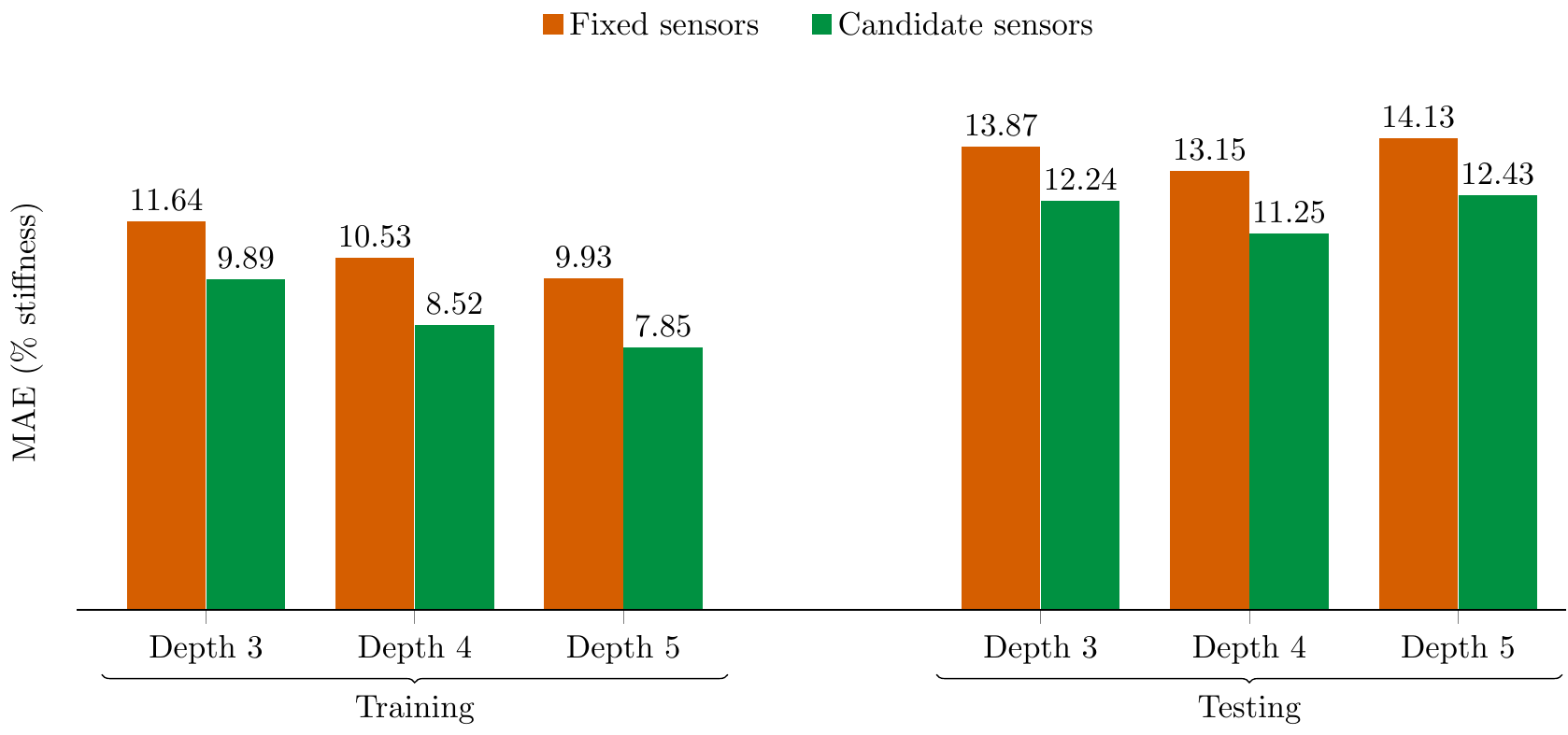}
\end{center}
\caption{Plot of classification performance (MAE) versus tree depth for optimal trees trained to classify $\mu_2$ using either the existing, fixed set of sensors, or the larger set of candidate sensor locations.}
\label{fig:sec04fixedcandidate}
\end{figure}
We see that the optimal trees determine that adding a subset of the candidate sensors would lead to improved classification performance. For example, in the depth three case, the optimal tree trained using the candidate set utilizes seven sensors that are already installed on the aircraft, and an additional six of the candidate sensors (see Figure \ref{fig:candidatesensors}). This set of sensors results in an MAE of $9.89$ as opposed to $11.64$ for the fixed sensor set. We observe similar trends for tree depths four and five. These results show how the optimal trees methodology can scale to a large number of sensors, and how utilizing the optimal trees methodology earlier in the digital twin development provides a principled approach to sensor placement that could lead to improved accuracy of the digital twin and fewer required sensors.
\subsection{Simulated self-aware UAV demonstration}\label{sec:UAVsim}
 %Simulation results putting it all together
 The results in the previous subsection show how optimal classification trees can be used to provide interpretable data-driven estimates of the two model parameters $\mu_1$ and $\mu_2$. In this subsection we present simulation results for an illustrative UAV scenario, which serves to demonstrate how a data-driven physics-based digital twin based on these optimal classification trees could be used to enable a self-aware UAV.

In this scenario, the UAV must fly safely through a set of obstacles to a goal location while accumulating structural damage or degradation. The UAV must choose either an aggressive flight path or a more conservative path around each obstacle. The aggressive path is faster, but requires the UAV to make sharp turns that subject the UAV to high structural loads (a 3g load factor). In contrast, the more conservative route is slower but subjects the UAV to lower structural loads (a 2g load factor). In pristine condition, the aircraft structure can safely withstand the higher 3g loading, but as the aircraft wing accumulates damage or degradation this may no longer be the case. Our self-aware UAV uses the rapidly updating digital twin in order to monitor its evolving structural state and dynamically estimate its flight capability. In particular, the physics-based structural model incorporated in the digital twin predicts that if the reduction in stiffness within either damage region exceeds 40\%, then a 3g load would likely result in structural failure. Thus, the UAV's control policy is to fall back to the more conservative 2g maneuver when the estimated reduction in stiffness exceeds this threshold. In this way the UAV is able to dynamically replan the mission as required in order to optimize the speed of the mission while avoiding structural failure.

We simulate a UAV path consisting of three obstacles and spanning 100 timesteps. To evaluate the UAV's decision-making ability, we simulate a linear reduction in stiffness in each of the damage regions from 0\% to 80\% over the 100 timesteps. At each timestep the UAV obtains noisy strain measurements from each of the 20 strain gauges. These measurements are used as inputs in the OCTs for estimating the parameters $\mu_1$, and $\mu_2$ (shown in Figures \ref{fig:OCT1} and \ref{fig:OCT2} respectively). The resulting parameter estimates are used to rapidly update the physics-based digital twin to be the corresponding model, $M_j\in\mathcal{M}$. This model in turn provides dynamic capability updates and informs the UAVs decision about which flight path to take. The results of this simulation are summarized in Fig.\ \ref{fig:video}, which shows snapshots at three different timesteps.\footnote{A video of the full simulation is available online at \href{https://kiwi.oden.utexas.edu/research/digital-twin}{https://kiwi.oden.utexas.edu/research/digital-twin}.}
\begin{figure}[h]
\begin{center}
\includegraphics[width=0.77\textwidth]{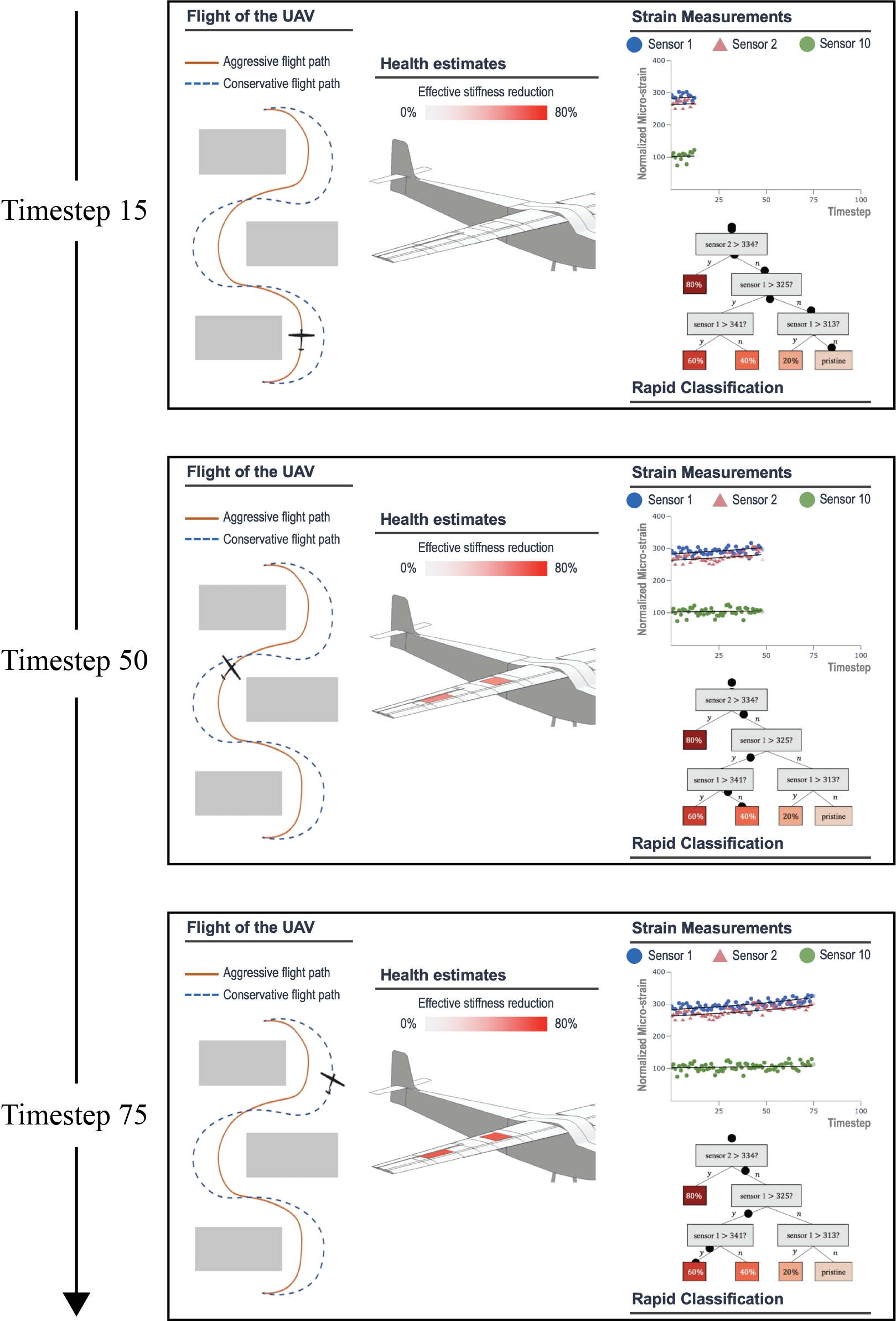}
\end{center}
\caption{Snapshots of the simulated UAV mission. Left: the UAV, obstacles, and possible flight paths. Center: UAV structural health estimates as provided by the digital twin. Right: Noisy load-normalized strain measurements (showing data from only three sensors) and the classification tree being used to classify the structural state of the UAV (showing here the OCT for parameter $\mu_1$). The first snapshot shows the UAV beginning in pristine condition and flying the aggressive flight path. In the second snapshot the digital twin estimates that the reduction in stiffness has progressed to 40\% in each damage region. At this point the UAV dynamically replans the mission, deciding to take the more conservative flight path in order to avoid structural failure. The final snapshot shows the UAV taking this conservative flight path around the third obstacle.}
\label{fig:video}
\end{figure}

\section{Conclusion}\label{sec:conclusion}
This work has developed an approach for enabling data-driven physics-based digital twins using interpretable machine learning. A library of physics-based models facilitates the generation of a rich dataset containing predictions of observed quantities for a wide range of asset states. This dataset is used to train an optimal classification tree that provides interpretable estimates of which model in the model library best matches observed data. Using this classifier online enables data-driven digital twin model adaptation. Our approach has been demonstrated using a case study in which a fixed-wing UAV uses structural sensors to detect damage or degradation on one of its wings. The sensor data is used in combination with optimal classification trees to rapidly estimate the best-fit model and update the digital twin of the UAV. The digital twin is then used to inform the UAV's decision making about whether to perform an aggressive maneuver or a more conservative one to avoid structural failure.

A limitation of our approach is that we have assumed that one has access to a model library that represents all the states a physical asset could be in. In practice this type of model library would be difficult to specify a-priori. For such cases, it would be beneficial to develop a principled method for detecting when the digital twin has gone beyond the limits of the model library. This could be done based on the uncertainty in the digital twin, or by comparing the error between the observed quantities predicted by the model and the observed data obtained from the asset. This detection could then trigger a manual inspection of the asset, and a subsequent enrichment of the model library to include the relevant state and reduce the uncertainty. Future work could explore variations in the optimization problem used to generate the classification trees. For example in some situations it would be beneficial to jointly optimize trees for different classification tasks, so that all trees can share a common sensor budget. Another area for future study is the robustness of optimal trees to sensor failures, and the investigation of whether the interpretability of the trees could enable one to detect and mitigate such failures.

\clearpage
\section*{Acknowledgments}
This work was supported in part by AFOSR grant FA9550-16-1-0108 under the Dynamic Data Driven Application Systems Program, by the SUTD-MIT International Design Center, and by The Boeing Company. The authors gratefully acknowledge C. Kays and other collaborators at Aurora Flight Sciences for construction of the physical UAV asset. The authors also acknowledge their collaborators D.J. Knezevic, D.B.P. Huynh, and M. Tran at Akselos for their work on the development of the reduced-order structural model of the UAV used in this work. Finally, the authors thank J. Dunn and others at Interpretable AI for the use of their software.

\bibliography{references}

\begin{thebibliography}{31}
\newcommand{\enquote}[1]{``#1''}
\providecommand{\natexlab}[1]{#1}
\providecommand{\url}[1]{\texttt{#1}}
\providecommand{\urlprefix}{URL }
\expandafter\ifx\csname urlstyle\endcsname\relax
  \providecommand{\doi}[1]{doi:\discretionary{}{}{}#1}\else
  \providecommand{\doi}{doi:\discretionary{}{}{}\begingroup
  \urlstyle{rm}\Url}\fi

\bibitem[{Hartmann and Van~der Auweraer(2020)}]{hartmann2020digital}
Hartmann, D., and Van~der Auweraer, H., \enquote{Digital Twins,} \emph{arXiv
  preprint arXiv:2001.09747}, 2020.

\bibitem[{Glaessgen and Stargel(2012)}]{glaessgen2012digital}
Glaessgen, E., and Stargel, D., \enquote{The Digital Twin Paradigm for future
  NASA and US Air Force Vehicles,} \emph{53rd AIAA/ASME/ASCE/AHS/ASC
  Structures, Structural Dynamics and Materials Conference 20th AIAA/ASME/AHS
  Adaptive Structures Conference 14th AIAA}, 2012, p. 1818.

\bibitem[{Li et~al.(2017)Li, Mahadevan, Ling, Choze, and Wang}]{li2017dynamic}
Li, C., Mahadevan, S., Ling, Y., Choze, S., and Wang, L., \enquote{Dynamic
  Bayesian Network for Aircraft Wing Health Monitoring Digital Twin,}
  \emph{AIAA Journal}, Vol.~55, No.~3, 2017, pp. 930--941.

\bibitem[{Tuegel et~al.(2011)Tuegel, Ingraffea, Eason, and
  Spottswood}]{tuegel2011reengineering}
Tuegel, E.~J., Ingraffea, A.~R., Eason, T.~G., and Spottswood, S.~M.,
  \enquote{Reengineering Aircraft Structural Life Prediction using a Digital
  Twin,} \emph{International Journal of Aerospace Engineering}, 2011.

\bibitem[{Kraft and Kuntzagk(2017)}]{kraft2017engine}
Kraft, J., and Kuntzagk, S., \enquote{Engine Fleet-Management: The Use of
  Digital Twins From a MRO Perspective,} \emph{ASME Turbo Expo 2017:
  Turbomachinery Technical Conference and Exposition}, American Society of
  Mechanical Engineers, 2017.

\bibitem[{Reifsnider and Majumdar(2013)}]{reifsnider2013multiphysics}
Reifsnider, K., and Majumdar, P., \enquote{Multiphysics stimulated simulation
  digital twin methods for fleet management,} \emph{54th AIAA/ASME/ASCE/AHS/ASC
  Structures, Structural Dynamics, and Materials Conference}, 2013, p. 1578.

\bibitem[{Jeon et~al.(2019)Jeon, Justin, and Mavris}]{jeon2019improving}
Jeon, H.~Y., Justin, C., and Mavris, D.~N., \enquote{Improving Prediction
  Capability of Quadcopter Through Digital Twin,} \emph{AIAA Scitech 2019
  Forum}, 2019, p. 1365.

\bibitem[{Kapteyn et~al.(2020)Kapteyn, Knezevic, Huynh, Tran, and
  Willcox}]{kapteyn2020digitalttwin}
Kapteyn, M.~G., Knezevic, D.~J., Huynh, D. B.~P., Tran, M., and Willcox, K.~E.,
  \enquote{Data-driven physics-based digital twins via a library of
  component-based reduced-order models,} \emph{Manuscript submitted for
  publication.}, 2020.

\bibitem[{Kennedy and O'Hagan(2001)}]{kennedy2001bayesian}
Kennedy, M.~C., and O'Hagan, A., \enquote{Bayesian calibration of computer
  models,} \emph{Journal of the Royal Statistical Society: Series B
  (Statistical Methodology)}, Vol.~63, No.~3, 2001, pp. 425--464.

\bibitem[{Zhang et~al.(2018)Zhang, De~Visser, and Chu}]{zhang2018aircraft}
Zhang, Y., De~Visser, C., and Chu, Q., \enquote{Aircraft damage identification
  and classification for database-driven online flight-envelope prediction,}
  \emph{Journal of Guidance, Control, and Dynamics}, Vol.~41, No.~2, 2018, pp.
  449--460.

\bibitem[{Dourado et~al.(2020)Dourado, Irmak, Viana, and
  Gordon}]{dourado2020bayesian}
Dourado, A.~D., Irmak, F., Viana, F., and Gordon, A., \enquote{Bayesian
  Calibration of Strain-Life Models with Priors from Similar Alloys,}
  \emph{AIAA Scitech 2020 Forum}, 2020, p. 2106.

\bibitem[{Zakrajsek and Mall(2017)}]{zakrajsek2017development}
Zakrajsek, A.~J., and Mall, S., \enquote{The development and use of a digital
  twin model for tire touchdown health monitoring,} \emph{58th
  AIAA/ASCE/AHS/ASC Structures, Structural Dynamics, and Materials Conference},
  2017, p. 0863.

\bibitem[{Zhao et~al.(2019)Zhao, Gupta, Regan, Miglani, Kapania, and
  Seiler}]{zhao2019component}
Zhao, W., Gupta, A., Regan, C.~D., Miglani, J., Kapania, R.~K., and Seiler,
  P.~J., \enquote{Component data assisted finite element model updating of
  composite flying-wing aircraft using multi-level optimization,}
  \emph{Aerospace Science and Technology}, Vol.~95, 2019, p. 105486.

\bibitem[{Chinesta et~al.(2018)Chinesta, Cueto, Abisset-Chavanne, Duval, and
  El~Khaldi}]{chinesta2018virtual}
Chinesta, F., Cueto, E., Abisset-Chavanne, E., Duval, J.~L., and El~Khaldi, F.,
  \enquote{Virtual, digital and hybrid twins: a new paradigm in data-based
  engineering and engineered data,} \emph{Archives of Computational Methods in
  Engineering}, 2018, pp. 1--30.

\bibitem[{Yucesan and Viana(2020)}]{yucesan2020hybrid}
Yucesan, Y.~A., and Viana, F., \enquote{A hybrid model for main bearing fatigue
  prognosis based on physics and machine learning,} \emph{AIAA Scitech 2020
  Forum}, 2020, p. 1412.

\bibitem[{Bertsimas and Dunn(2017)}]{OCT}
Bertsimas, D., and Dunn, J., \enquote{Optimal classification trees,}
  \emph{Machine Learning}, Vol. 106, No.~7, 2017, pp. 1039--1082.

\bibitem[{Bertsimas and Dunn(2019)}]{bertsimas2019machine}
Bertsimas, D., and Dunn, J., \emph{Machine learning under a modern optimization
  lens}, Dynamic Ideas LLC, 2019.

\bibitem[{Allaire et~al.(2012)Allaire, Biros, Chambers, Ghattas, Kordonowy, and
  Willcox}]{allaire2012dynamic}
Allaire, D., Biros, G., Chambers, J., Ghattas, O., Kordonowy, D., and Willcox,
  K., \enquote{Dynamic Data Driven Methods for Self-aware Aerospace Vehicles,}
  \emph{Procedia Computer Science}, Vol.~9, 2012, pp. 1206--1210.

\bibitem[{Lecerf et~al.(2015)Lecerf, Allaire, and
  Willcox}]{lecerf2015methodology}
Lecerf, M., Allaire, D., and Willcox, K., \enquote{Methodology for dynamic
  data-driven online flight capability estimation,} \emph{AIAA Journal},
  Vol.~53, No.~10, 2015, pp. 3073--3087.

\bibitem[{Singh and Willcox(2017)}]{singh2017methodology}
Singh, V., and Willcox, K.~E., \enquote{Methodology for path planning with
  dynamic data-driven flight capability estimation,} \emph{AIAA Journal}, 2017,
  pp. 2727--2738.

\bibitem[{Fisher(1936)}]{fisher1936use}
Fisher, R.~A., \enquote{The use of multiple measurements in taxonomic
  problems,} \emph{Annals of eugenics}, Vol.~7, No.~2, 1936, pp. 179--188.

\bibitem[{Dua and Graff(2017)}]{Dua:2019}
Dua, D., and Graff, C., \enquote{{UCI} Machine Learning Repository,} , 2017.
\newblock \urlprefix\url{http://archive.ics.uci.edu/ml}.

\bibitem[{Breiman(2017)}]{CART}
Breiman, L., \emph{Classification and regression trees}, Routledge, 2017.

\bibitem[{Murthy and Salzberg(1995)}]{murthy1995lookahead}
Murthy, S., and Salzberg, S., \enquote{Lookahead and pathology in decision tree
  induction,} \emph{IJCAI}, Citeseer, 1995, pp. 1025--1033.

\bibitem[{Heath et~al.(1993)Heath, Kasif, and Salzberg}]{SADT}
Heath, D., Kasif, S., and Salzberg, S., \enquote{Induction of oblique decision
  trees,} \emph{IJCAI}, Vol. 1993, 1993, pp. 1002--1007.

\bibitem[{Murthy et~al.(1994)Murthy, Kasif, and Salzberg}]{OC1}
Murthy, S.~K., Kasif, S., and Salzberg, S., \enquote{A system for induction of
  oblique decision trees,} \emph{Journal of Artificial Intelligence Research},
  Vol.~2, 1994, pp. 1--32.

\bibitem[{Breiman(2001)}]{rf}
Breiman, L., \enquote{Random Forests,} \emph{Machine learning}, Vol.~45, No.~1,
  2001, pp. 5--32.

\bibitem[{Friedman(2001)}]{xgboost}
Friedman, J.~H., \enquote{Greedy function approximation: a gradient boosting
  machine,} \emph{Annals of statistics}, 2001, pp. 1189--1232.

\bibitem[{"Gurobi~Optimization(2019)}]{gurobi}
"Gurobi~Optimization, L., \enquote{"Gurobi Optimizer Reference Manual",} ,
  2019.
\newblock \urlprefix\url{"http://www.gurobi.com"}.

\bibitem[{Drela(1999)}]{drela1999integrated}
Drela, M., \enquote{Integrated simulation model for preliminary aerodynamic,
  structural, and control-law design of aircraft,} \emph{40th Structures,
  Structural Dynamics, and Materials Conference and Exhibit}, 1999, p. 1394.

\bibitem[{{Interpretable AI, LLC}(2020)}]{InterpretableAI}
{Interpretable AI, LLC}, \enquote{Interpretable AI Documentation,} version
  1.1.0, 2020.
\newblock \urlprefix\url{https://www.interpretable.ai}.

\end{thebibliography}
\end{document}